\documentclass[journal,comsoc]{IEEEtran}
\usepackage{cite}
\usepackage{amsmath,amssymb,amsfonts}
\usepackage{subcaption}

\usepackage{fancyhdr}
\usepackage{algorithmic}
\usepackage{graphicx}
\usepackage{textcomp}
\def\BibTeX{{\rm B\kern-.05em{\sc i\kern-.025em b}\kern-.08em
    T\kern-.1667em\lower.7ex\hbox{E}\kern-.125emX}}
\begin{document}
\chead{This work has been submitted to the IEEE for possible publication. Copyright may be transferred without notice, after which this version may no longer be accessible.
}

\title{BEAT: Blockchain-Enabled Accountable and Transparent Infrastructure Sharing in 6G and Beyond}
\author{\IEEEauthorblockN{Tooba Faisal,
Mischa Dohler,
Simone Mangiante and 
Diego R. Lopez}}
\maketitle
\thispagestyle{fancy}

\begin{abstract}
It is widely expected that future networks of 6G and beyond will substantially improve on 5G. Technologies such as Internet of Skills and Industry 4.0 will become stable and viable, as a direct consequence of networks that offer sustained and reliable mobile performance levels.

The primary challenges for future technologies are not just low-latency and high-bandwidth. The more critical problem Mobile Service Providers (MSPs) will face will be in balancing the inflated demands of network connections and customers' trust in the network service. That is, being able to interconnect billions of unique devices while adhering to the agreed terms of Service Level Agreements (SLAs). To meet these targets, it is self-evident that MSPs cannot operate in a solitary environment. They must enable cooperation among themselves in a manner that ensures trust, both between themselves as well as with customers.

In this study, we present the BEAT (\textbf{B}lockchain-\textbf{E}nabled \textbf{A}ccountable and \textbf{T}ransparent) Infrastructure Sharing architecture. BEAT exploits the inherent properties of permissioned type of distributed ledger technology (i.e., permissioned distributed ledgers) to deliver on accountability and transparency metrics whenever  infrastructure needs to be shared between providers.  We also propose a lightweight method that enables device-level accountability. BEAT has been designed to be deployable directly as only minor software upgrades to network devices such as routers. Our simulations on a resource-limited device show that BEAT adds only a few seconds of overhead processing time -- with the latest state-of-the-art network devices, we can reasonably anticipate much lower overheads.

\end{abstract}


\maketitle

\section{Introduction}
The introduction of 5G has launched the era of low latency and high bandwidth applications; concepts such as the Internet of Skills and Industry 4.0 provide us with both a goal and a vision of the future of telecommunications technologies.  Indeed,  these applications hold the potential to contribute substantially to the prosperity of humankind. For instance, the Internet of Skills can allow for knowledge and expertise developed for the treatment of highly contagious viruses, like the Ebola in~\cite{dohler2017internet} to be applied in real-time to the treatment of today's Covid-19 pandemic. Using robotics, medical staff can treat the covid patients remotely without risking their own lives. Similarly, Industry 4.0 will enable the demands of the growing world population to be met with the production and supply of goods with maximal, and perhaps optimal automation.

For the applications such as automated remote surgery, enabled by the Internet of Skills~\cite{dohler2017internet}, there is the inherent mandate for network stability and guaranteed low latency~\cite{6G_governance}. These critical requirements are above and beyond those expected of  so-called ``non-essential'' applications, such as listening to music or watching a movie; for which today's ``best-effort'' Internet delivery systems suffice admirably. Particularly where human lives are at stake, guaranteed performance levels provide a compelling reason for us to engineer network systems that can meet these requirements.


In this work, we advocate two key directions to increase the network capability and enable scalability in the future networks of 6G and beyond. Firstly, one realistic solution to coping with the growing demand is the \emph{Mobile Service Providers'~(MSP) collaboration}. Indeed, we have already witnessed, for instance, the recent  merger between Virgin Media and O2~\cite{o2_vm}, which will make Virgin Media one of the biggest provider and the planned masts' sharing agreement between  O2, Three and Vodafone to boost the rural coverage~\cite{sharing2}. In another case, Japanese telecom operator KDDI and SoftBank have announced a sharing agreement of RAN sharing through Multi-operator RAN~(MORAN) to provide 5G network coverage to rural areas. In a similar deal, in 2019, China Telecom and China Unicom started to share their 5G infrastructure in a deal that allows them to save approximately \$13.2 billion~\cite{5Gjapan_china_deals}. 

Clearly, such resource sharing agreements are already part of MSPs' operating practices and are proving to be both financially and technically viable; however, we note that such agreements typically  involve only a limited number of infrastructure providers bound by a long-term contract model. As 5G transitions inevitably towards 6G,  we anticipate that agreements are likely to become both more compelling and common.

Secondly, millions of new devices, such as connected cars and  industrial IoT, are now connecting to mobile networks on a daily basis. 6G must be ready to meet the scale and demand of not just millions, but billions of network devices -- according to~\cite{number_IoT_non_IoT} the number of IoT will increase to 30.9 billion and non-IoT to 10.3 billion by 2025. Paradoxically, one of the obstacles to meeting the inflated demand is the device supply -- the vendors must scale up their production capacity to produce the equipment that can meet the future user demand for network devices. However, from the perspective of network engineering, in the case of a production bottleneck, the network services should not be impacted at all. 

A realistic approach to balancing network scalability and device vendors' workloads is to enable \emph{multi-vendor} equipment. The current deployment of 5G, indeed, enables a multi-vendor environment; for instance, O-RAN allows the disaggregation of hardware and software of RAN and enables a multi-vendor environment. Yet, the openness is cumbersome and complicated process owing to various reasons such as network management, interoperability and even geopolitical considerations. In a recent report~\cite{oran_project_report}, ``Lack of Single Accountable Supplier'' is listed as one of the challenges for OpenRAN viability.

Having established these two key motivations, we note that both techniques have challenges and requirements. While one can argue for the need for trust among the MSPs, one can also argue about the dangers and accountability challenges in a multi-vendor environment. We believe that the network sharing systems in 6G and beyond are required to be \emph{transparent} such that all the stakeholders can \emph{monitor} and manage the SLA\emph{transparently}. In case of violation, a party at fault can be held accountable, and all the penalties should be applied and paid for \emph{automatically} without any delay.

\begin{figure}
    \centering
    \includegraphics[width=80mm]{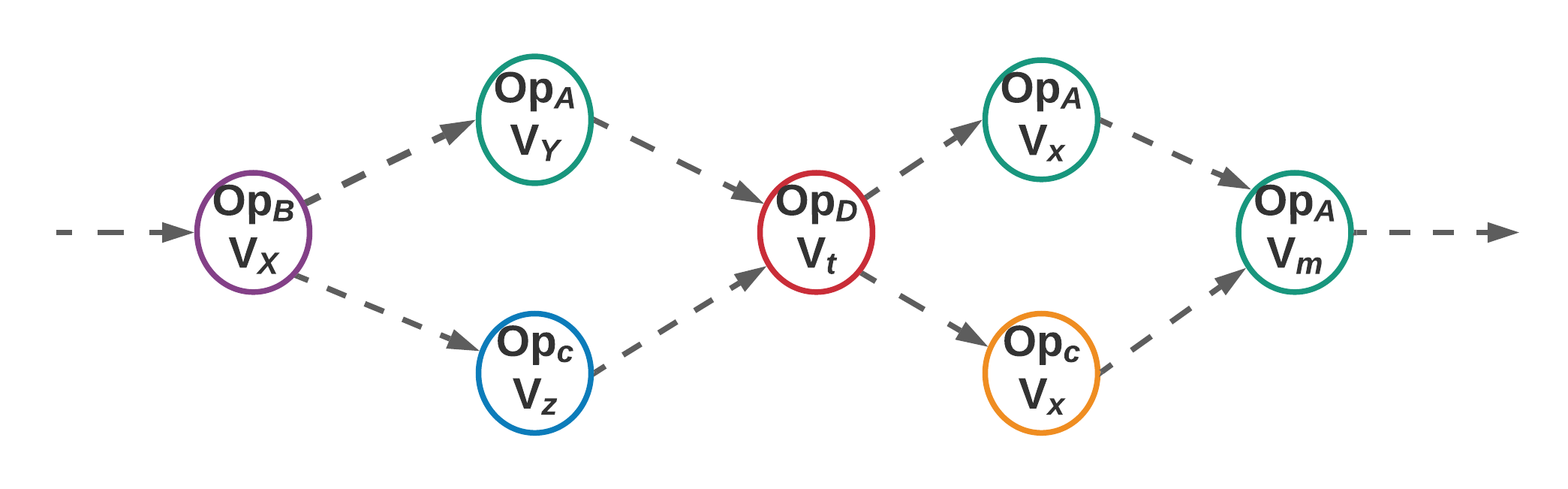}
    \caption{BEAT Vision -- We envision a multi-operator and multi-vendor network infrastructure. Network infrastructure is formed by multiple operators (e.g., $Op_A$,$Op_B$,$Op_C$,$Op_D$) and devices are provided by multiple vendors(e.g., $V_m$,$V_x$,$V_y$,$V_z$))}
    \label{fig:beat_overview}
\end{figure}


Network sharing mechanisms proposed by the research community tend to  focus primarily on the efficiency of resource sharing issues; for instance \cite{crippa2017resource} discusses the efficient resource sharing in network slices, but not accountable resource reservation at the device level. Samdani et al.\cite{samdanis2016network} rely on a centralised entity for network sharing but do not  provide any mechanism to maintain records for future audits nor do they provide transparency to the network users for the services provided.

To this end, this paper introduces our proposed solution to the aforementioned problems and outlines a viable way forward for the industry. We introduce a \textbf{B}lockchain-\textbf{E}nabled \textbf{A}ccountable and \textbf{T}ransparent (BEAT) Infrastructure sharing through smart contracts residing on Permissioned Distributed Ledgers (PDLs). PDLs are a particular type of distributed ledger that works between a closed group of mutually non-trusted parties. Access is managed via stringent access control mechanisms, that is,~only authorised members can access a PDL, making them ideal for business-like applications. Furthermore, PDLs are immutable and contain executable smart contract codes which gets deployed and executed on the ledger. 

The required SLAs are installed as smart contracts on a PDL, whereas the PDL nodes themselves are installed  on the network devices, either bare-metal or in a virtualised environment. They record the infrastructure usage by monitoring  packet flows and by recording the relevant data to the PDL. The high degree of automation in this design can cope with surges in  demand for network services and make the infrastructure sharing plug-and-play without a long contractual process.

The remainder of the paper is organised as follows. A review of existing and related work is conducted in Section~\ref{sec:related_work}. In Section~\ref{sec:sys_arch2}, we present "BEAT", our blockchain-based end-to-end architecture. We present an accountability-enabled protocol ``Interrogation-Protocol'' in Section~\ref{sec:int_proto} and evaluate our work in Section~\ref{sec:evaluation} by addressing the questions of the resource and performance overhead of PDLs. Possible future research directions are presented in Section~\ref{sec:future_work}. The paper ends with the concluding remarks in~Section~\ref{sec:conclusion}.

\section{Related Work}
\label{sec:related_work}
The use of blockchain-based distributed ledger technologies for network resource sharing is an emerging area of research that has been considered in several recent studies.

We restrict our discussion here to the results that are most closely related to our own efforts and,  in particular, focus on resource sharing with distributed ledger technologies and accountability challenges.

\subsection{Resource Sharing}
The concept of infrastructure and spectrum sharing in 5G and beyond with blockchain is presented by ~\cite{maksymyuk2020blockchain} and \cite{maksymyuk2019blockchain}. 
Maksymyuk et al. in~\cite{maksymyuk2019blockchain} also propose a coalition algorithm for unused spectrum sharing. Maksymyuk et al. in \cite{maksymyuk2020blockchain} further list the tokenisation model for spectrum, infrastructure, and service pricing. They also provide a dynamic and smart contract focused protocol for spectrum sharing. However, both studies focus on providing a broad vision of the possibilities of exploiting blockchain technology for future networks; none of them provides a detailed, device-level architecture for infrastructure sharing.

Another Distributed Ledger Technology (DLT)-focused resource reservation work is \emph{Blockchain Network Slice Broker}~\cite{backman2017blockchain} and based on the \emph{Network Slice Broker} of \cite{samdanis2016network}. In that work, tenants (such as Over-the-Top providers) can request network services from the Mobile Network Operators~(MNOs) on-the-fly. The SLAs for  allocation are recorded  to a distributed ledger through smart contracts. However, the actual resource usage at a device is not recorded, and the problem of accountability in network sharing is not addressed.~The extension of Network Slice Broker~(NSB)~\cite{samdanis2016network} is presented in~\cite{NSB_chain} and provides a blockchain-focused architecture for network slice auctions, in which infrastructure providers allocate network slices through an intermediate entity intermediate broker which further allocates resources to tenants. However, the NSB does not discuss the problem of accountability.

A transparent on-the-fly Software Defined Network~(SDN) based technique for radio resource sharing architecture is proposed by~\cite{jiang2017radio}. In this work, a customer can connect to any available operator where the resources are available and is dependent on a third-party entity~(essentially an SDN-Server), which keeps track of the available resources throughout the participating service providers. However, this work is limited to resource provisioning and does not discuss the prospects of low-level~(i.e., switch level) sharing or accountability due to SLA violation.


The inter-operator network sharing architecture, specifically for smaller/denser cells, is proposed in~\cite{okon2020blockchain} and in \cite{mafakheri2018blockchain}. The network sharing SLAs between MNOs are stored as smart contracts on a distributed ledger and executed with service requests through an SDN layer.~A blockchain-focused unlicensed spectrum sharing approach is presented by~\cite{maksymyuk2019blockchain}. Although this work discusses the prospect of network sharing among different players, it does not provide any architecture of network sharing; however it provides a game-theoretic algorithm for unlicensed spectrum sharing.
\subsection{Accountability}
Accountability is the fundamental property of BEAT, presented here as a novel architecture, that should be scaled up to future as-yet undefined 6G technologies.~Network users should adhere to the SLA and not misbehave with other network users. The Interrogation Protocol (Section~\ref{sec:int_proto}) proposes a solution to enable accountability in resource allocation. The devices (e.g., routers and switches) are stitched together to enable resource sharing. As this idea is similar to the Internet network model, we study and take inspiration from accountability in the context of Internet Service Providers and the Internet.

We note that solutions to enable accountability at the Administrative-Domain(AD) level have previously been  presented in AudIt \cite{argyraki2007loss} -- a network traffic auditing protocol. Similar to BEAT, in AudIt, the service providers report their QoS parameters, but at the Autonomous System (AS) granularity, rather than at device granularity, as in our approach. In addition, the reports collected are not recorded by all the devices immutably for a future audit. Similarly, we find that the FAIR (Forwarding Accountability for Internet Reputability) -- accountability focused protocol of~\cite{pappas2015fair}, enables the AS-level accountability through packet inscription. This is in contrast to the BEAT solution, in which accountability is enabled at the device level without any packet modifications.

\section{Distributed Ledgers and Smart Contracts Background}
\label{sec:dlt_background}

Distributed Ledger Technology is a broad term and its applications include distributed ledgers and smart contracts.
Several definitions of distributed ledgers are presented in the literature~\cite{rauchs2018distributed}. In this work we define Distributed Ledgers as:
\textit{A network of distributed devices, which uses methods and techniques to replicate data across multiple devices within a distributed network to ensure data integrity and transparency.}

Distributed ledgers are  data structures in which data with a unique cryptographic signature and a timestamp is replicated across different machines across the network.  When a node sends a transaction request, that is, a request to write certain data in the ledger, based on the consensus algorithm chosen for the distributed ledgers, the transaction is accepted or rejected.  When several nodes send data to a shared/distributed ledger, it is difficult to verify the authenticity of data. That is, if a node is malicious and cheats (e.g., recording incorrect data to the ledger), all the nodes will receive the wrong data and the whole ledger will be compromised. To this end, distributed ledgers add the data to the ledger only after a consensus is reached. Nevertheless, Distributed ledgers do suffer from known issues and  limitations such as serialisation (in some distributed ledger types), out-of-sync data and fraud~\cite{dlt_basics}. 




\paragraph{Blockchain}

The typical problem of serialisation in distributed ledger is solved in Blockchains. A blockchain or ``Block-chain'' is  \textit{a type of distributed ledger which is formed through verified and serialised blocks replicated across the network nodes}.  That is, the members of the blockchain agree to the consensus of block generation. Blocks may be generated, for example, over a fixed time duration and/or by solving a cryptographic puzzle; once generated, they are added to the blockchain through consensus. The concept of blockchain dates back to 1991~\cite{tijan2019blockchain} when Haber and Stornetta presented the idea of time stamping and linking documents such that no party can amend the records~\cite{blockchain_origin_1991}.

In literature, distributed ledgers and blockchain are sometimes used interchangeably~\cite{rauchs2018distributed}. Distributed ledgers is a broad term, and indeed blockchain is a particular type of distributed ledger, but distributed ledgers do not necessarily need to rely blocks.

We may further classify distributed ledgers into two main categories:
\begin{itemize}
\item Permissionless distributed ledgers
\item Permissioned distributed ledgers
\end{itemize}
Note that any distributed ledger which generates serialised blocks of data, can be classed as a blockchain. This condition stands for both the permissionless and permissioned scenario discussed below:

\subsubsection{Permissionless Distributed Ledgers}
In permissionless distributed ledgers, the ledger participants can join freely without any access-control mechanisms. The best-known examples of permissionless distributed ledgers are Bitcoin and Ethereum’s Mainnet. Since participants can join the network without any checks, permissionless distributed ledgers are often not suitable for business-like applications~\cite{wust2018you} and they are at higher risk of malicious attacks, as well as having limited scalability.

\subsubsection{Permissioned Distributed Ledgers (PDLs)}
Permissioned distributed ledgers are a special type of distributed ledgers, formed by a consortium of members. Generally, these members are somehow known to each other and members are allowed in the network with strict control mechanisms. Permissioned distributed ledgers are also known as Consortium Blockchains. The best-known examples are the Hyperledger Fabric and R3 Corda. Due to the security advantages conferred by controlled access, they are suitable for business-like applications and are receiving increased attention from the industry and standardisation bodies. For instance, ETSI ISG (Industry Specification Group) PDL is focused on developing standards for industries, specific to permissioned distributed ledgers.

 A brief comparison of permissionless and permissioned distributed ledgers is given in Table~\ref{tab:dlt_comparison}.

\begin{table}[h!]
\centering
 \begin{tabular}{|p{0.25\linewidth}|p{0.25\linewidth}|p{0.25\linewidth}|}
\hline 
  Property &  Permissionless DL &Permissioned DL\\ [0.5ex] 
 \hline\hline
 Scalability & Low &Medium to High  \\[0.5ex]
 Eternity &  High&Governance-dependent \\[0.5ex]
 Latency~\cite{wust2018you}  & High& Low\\[0.5ex]
 Throughput~\cite{wust2018you} 	& Low& High\\[0.5ex]
 Control & No Control& Governance Controlled \\[0.5ex]
 Access Control  & No& Yes\\[0.5ex]

 \hline
 \end{tabular}
 \caption{Comparison of Permissionless and Permissioned Distributed Ledgers}
 \label{tab:dlt_comparison}
\end{table}

\subsection{Smart Contracts}
Distributed ledgers (both permissioned and permissionless) on their own are static. That is, they record the data as per the consensus to the ledger and do not perform executions. In order that users may run executable code, however, distributed ledgers rely on smart contracts, which are snippets of code installed on the distributed ledger itself. Since they are installed on the ledger, smart contracts inherit the properties of distributed ledgers.

Some of the properties of a smart contract are as follows:

\paragraph{Transparency}
Typically, all the nodes of a distributed ledger keep an identical copy of the ledger. Therefore, a smart contract is also replicated on every node of the ledger,  and any changes to the contract become transparent to all the members of the system.
\paragraph{Monitorability}
It follows from transparency that the execution of a smart contract is replicated throughout the ledger/nodes.
\paragraph{Auto-execution}
Smart contracts are executable autonomously with certain pre-defined conditions.

\paragraph{Immutability}
As distributed ledgers are immutable, it follows that smart contracts are also immutable -- once installed, a smart contract cannot be changed or deleted.

 

\section{Proposed System Architecture}
\label{sec:sys_arch2}

\begin{figure*}

  \includegraphics[width=\textwidth]{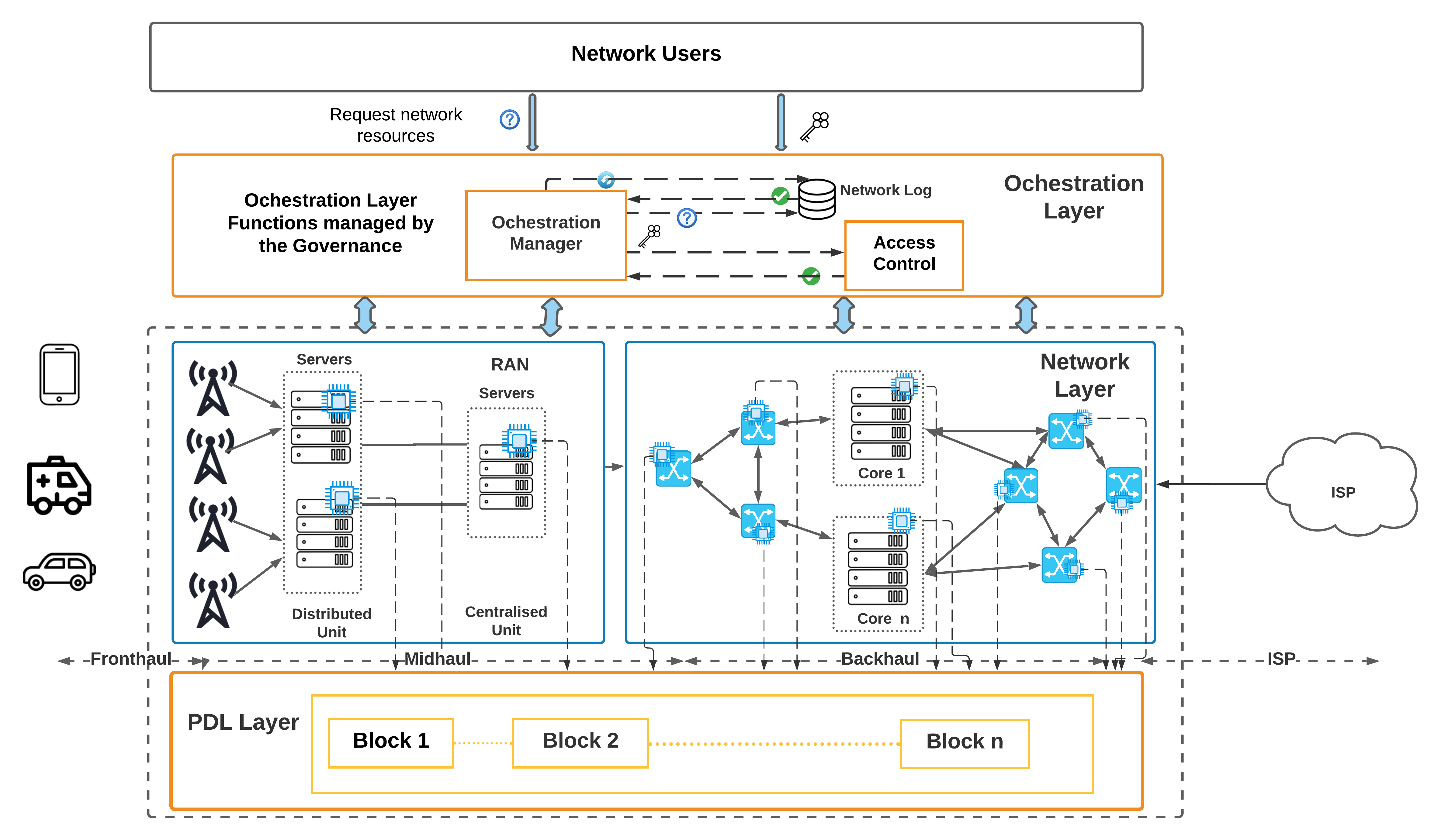}
  \caption{Network resource sharing architecture }
  \label{fig:system_arch_1}
\end{figure*}

BEAT is an automated architectural solution with three key elements: (1) resource sharing with (2) distributed ledgers and smart contracts that enable (3) accountability and transparency, with these objectives being achieved through PDLs and smart contracts. In this section, we explain the BEAT architecture in detail.

\subsection{BEAT's Architecture}
\label{subsec:beat_arch}
The architecture is underpinned by the following actors:
\begin{itemize}
    \item \textbf{Network Users:} Who can be further classified into 1) \textit{Network Owners} -- A party or group of participants who own the infrastructure, or 2) \textit{Network Tenants} -- the party who lease the infrastructure from network owners, or 3) \textit{both }--  own some of the network infrastructure, which other tenants can lease. Moreover, they also  lease/rent some network infrastructure from the owners to serve their customers.
    
    \item \textbf{Device Vendors:} They provide network devices~(e.g. routers and switches). 
    \item \textbf{Governance:} Network governance is a decision-making committee; formed by the consortium of the network users and include their representatives.  It is up to the network users to decide the strategy (e.g., through voting) by which governance representatives are chosen. Governance takes management decisions, such as access control and dispute resolution. Governance members may have control over their own devices only and do not control their peers' devices. Therefore, in the situations of a malicious user in the governance, only one vote will be compromised. 

\end{itemize}

Typically, it is a challenge in distributed ledgers to \emph{truly} distribute the power among the nodes; some entities somehow try to control the network~\cite{ZACHARIADIS2019105}. In BEAT, every network user will have equal representation to vote and nominate their representatives, regardless of their stake in the network infrastructure. Our governance model is similar to the Network Administrative Organisation (NAO) of ~\cite{provan2008modes}. In NAO, a third party manages the network rather than the network members. In BEAT, the governance includes the PDL members, rather than an external entity.

In addition to these actors, BEAT's architecture comprises three operational layers, namely the Orchestration, Network and PDL layers. These are illustrated in Figure~\ref{fig:system_arch_1} and we discuss them next.


\subsection{Orchestration Layer}
\label{subsec:orch_layer}

The \textbf{Orchestration Layer} is the top layer and handles the network resource requests from the tenants. Its operations are similar to the ETSI's Management And Orchestration Layer~(MANO). This layer is maintained and managed by the governance of the PDL. It oversees the network operations and allocation decisions such as setting up network access, lease duration, price, and privileges. The Orchestration Layer is managed by the governance of the network.

Shown in Figure~\ref{fig:orch_process}, the Orchestration Layer has three main components:
\begin{figure}
    \centering
    \includegraphics[width=0.49\textwidth]{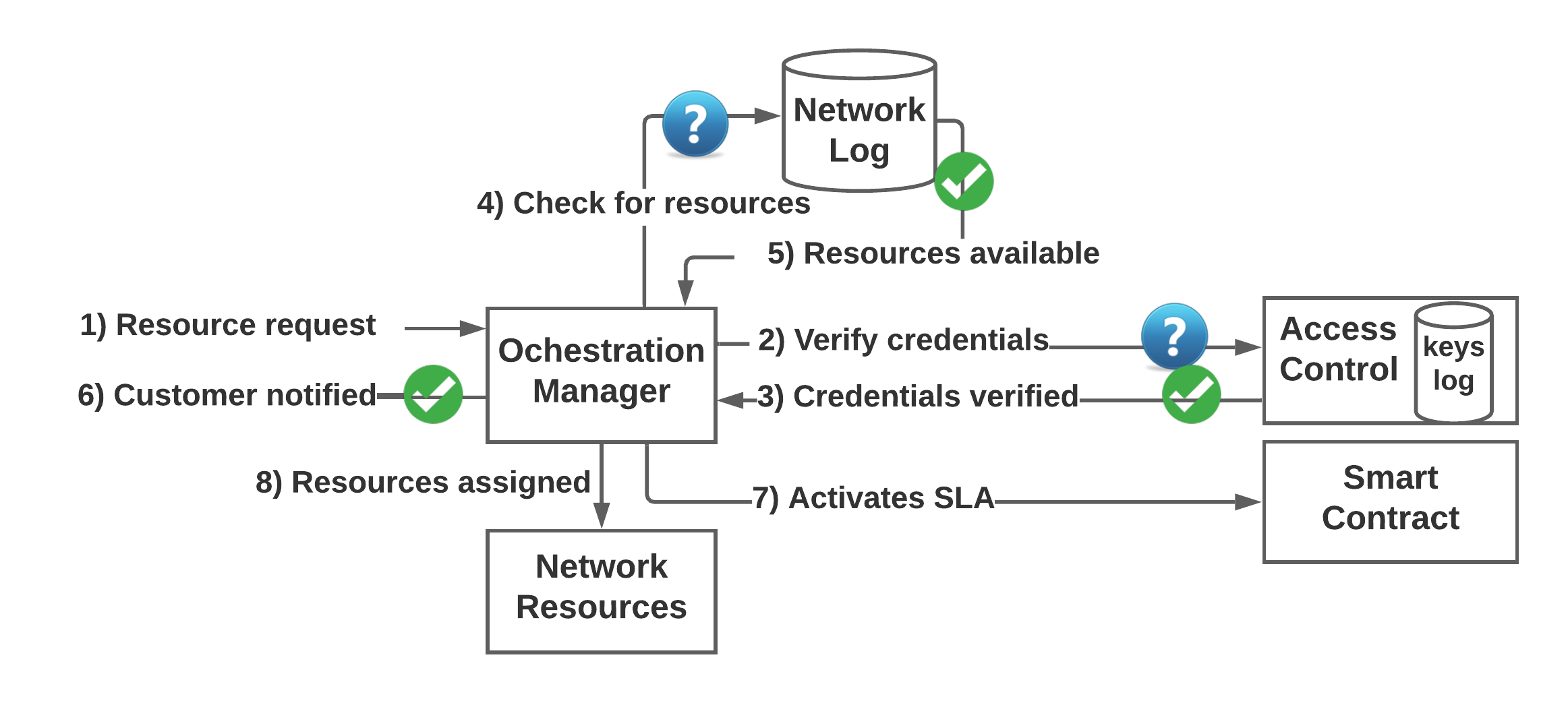}
    \caption{BEAT Orchestration Process}
    \label{fig:orch_process}
\end{figure}

\begin{enumerate}
    \item \textbf{Orchestration Manager:} This serves the incoming requests and has  features such as \emph{Universal View} similar to the Software-Defined Mobile Network Orchestrator~(SDM-O) discussed in~\cite{crippa2017resource} and the  SDN-Server of~\cite{jiang2017radio}. When a network participant joins the network, the Orchestration Manager assigns the credentials and keeps the record. The Orchestration Manager also allocates a node ID to a device. This node ID is different from the Layer 2/3 addresses because the devices may change their IP addresses anytime whereas the node ID must remain the same.
    \item \textbf{Access Control:} An access control verification entity which maintains a database to keep the record of credentials and replies to  access control confirmation queries from the Orchestration Manager.
    \item \textbf{Network Log:}  A database to maintain network resource logs.
\end{enumerate}


The Orchestration Manager allocates the network resources through strict access controls. The first function of the Orchestration Manager is to install network SLAs as smart contracts on the PDL. Two different types of SLAs are installed on the PDL, 1) Resource Orchestration SLA and 2) QoS Monitoring SLA.

The \emph{Resource Orchestration SLA (RO-SLA)} -- is the SLA which is the initial agreement between the network users (e.g., an owner and tenant) and is executed at the start of the service. The governance installs RO-SLA at the earliest (e.g., at the time of PDL formation) and executes it with every network request. RO-SLA maintains the resource allocation details, such as  route identity and QoS parameters agreed.

The \emph{QoS Monitoring SLA~(QM-SLA)} -- this is an SLA for  quality monitoring  which records the per flow data  to the PDL.

\subsection{Network and PDL Layers}
Infrastructure and network resources, such as switches and routers, form the \textbf{Network Layer} of BEAT; these are managed and maintained by its Orchestration Layer.  Network devices have a maximum threshold capacity to forward network traffic while maintaining agreed service levels. As service quality would get affected beyond these limits, network access must be monitored and controlled.

When the tenants request network resources, the Orchestration Manager will verify from the access control entity if the tenant has an agreement in place already. If an agreement is present, the Orchestration Manager will then check the status of the network from the \emph{Network Log}, namely the current load on each path of the network and on each device. If the network has resources available, the Orchestration Manager will send a confirmation message to the tenant. Next, a smart contract will be executed to initialize the SLA and then orchestrate the network resource for the tenant. In the event that the capacity is not available, the tenant can simply wait for the resources to become available. 

BEAT advocates a multi-operator and multi-vendor environment. Therefore, stakeholders must know the performance and usage of the network components at a very fine-grained level. Such performance metrics are required for future SLA compliance and accountability of the sharing agreement.

 The infrastructure usage in BEAT is recorded at the device level and transparently shared with smart contracts at the \textbf{PDL Layer}. Every device in BEAT is equipped with a PDL node and can execute smart contracts to record relevant data to the PDL. At the macro-level, all the devices together form a PDL within the network infrastructure.

\subsection{Automated Recording with Smart Contracts}
\label{subsec:auto_sc}
A tenant's main objective is to get agreed on service levels to meet its customers' demands. Service quality can get affected for several reasons, such as a slower path or network device malfunctioning. In the event of service degradation, the tenant is entitled to get compensation if applicable and without any hassle. In such cases, all the stakeholders (i.e., network operators and vendors) would blame each other to avoid paying the penalty. Therefore, there should be a mechanism to record the service data, which no party can deny. 

In BEAT, the flows' data is recorded to the PDL through smart contracts. For each flow, both the source and  destination  records the relevant data to the PDL. PDLs are immutable, which means that the data recorded by them cannot be deleted. Moreover, PDLs are transparent, and all the participants for the consortium can see the flow source and destination information in the PDL. To resolve these two problems of scalability and privacy, we install minimum hashed data per flow to the PDL; specifically, \emph{node ID}, \emph{source IP address}, \emph{destination IP address} and \emph{timestamp}. For packet $i$ and node $j$ will record the following data $D$ to the smart contract:
\begin{equation}
     D_ij = SHA3(node\_ID_j, src\_IP_i , dst\_IP_i,ts_i). 
\label{eq:data}
\end{equation}

\begin{figure}
    \centering
    \includegraphics[width=0.49\textwidth]{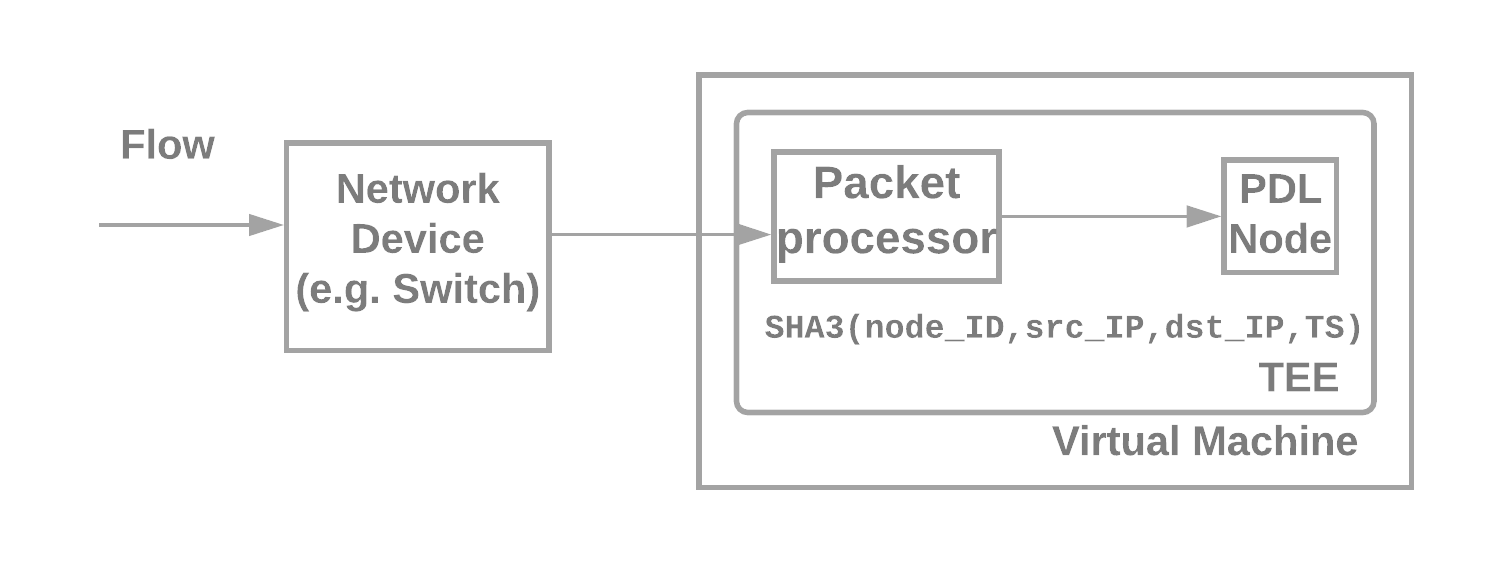}
    \caption{The Packet Processor -- records the packet data (i.e., \emph{node\_ID, src\_IP, dst\_IP, timestamp}) to the PDL node }
    \label{fig:packet_processor}
\end{figure}

\subsection{Trusted Execution Environments for Flow Monitoring}
\label{subsec:TEE}

The data is recorded to the PDL by means of a smart contract that is executed by the \emph{Packet Processor} -- which is a software script that runs on a virtual machine within the PDL node adjacent to the device; see Figure~\ref{fig:packet_processor}.  The packet processor extracts the data from the flow, hashes it and executes a smart contract to record the data to the PDL node. In BEAT, the owner has no control over this virtual machine, which thereby enables secure and trustworthy data recording. On the other hand, both the Packet Processor and the PDL node are installed on operator and/or vendor-controlled devices, whom tenants may not trust them as owners could tamper with data. Some device owners may, in addition to this, behave maliciously -- for example, they may intentionally drop a packet and claim that the packet was sent from the source and was dropped in the network. It is difficult to dispute such claims due to the best-effort nature of today’s internet. 

To circumvent these issues, BEAT adds another layer of security and wraps the Packet Processor and the PDL node inside a Trusted Execution Environment (TEE).  It is to be noted here that TEE is a separate secure processing system that solves this trust issue. The governance can give the tenants controlled access to the virtual machine -- this access is decided by the PDL voting mechanisms and will rotate among the tenants to enable trustworthy record keeping.

\section{Interrogation Protocol}
\label{sec:int_proto}

\begin{figure}[!h]
    \centering
    \includegraphics[width=0.48\textwidth]{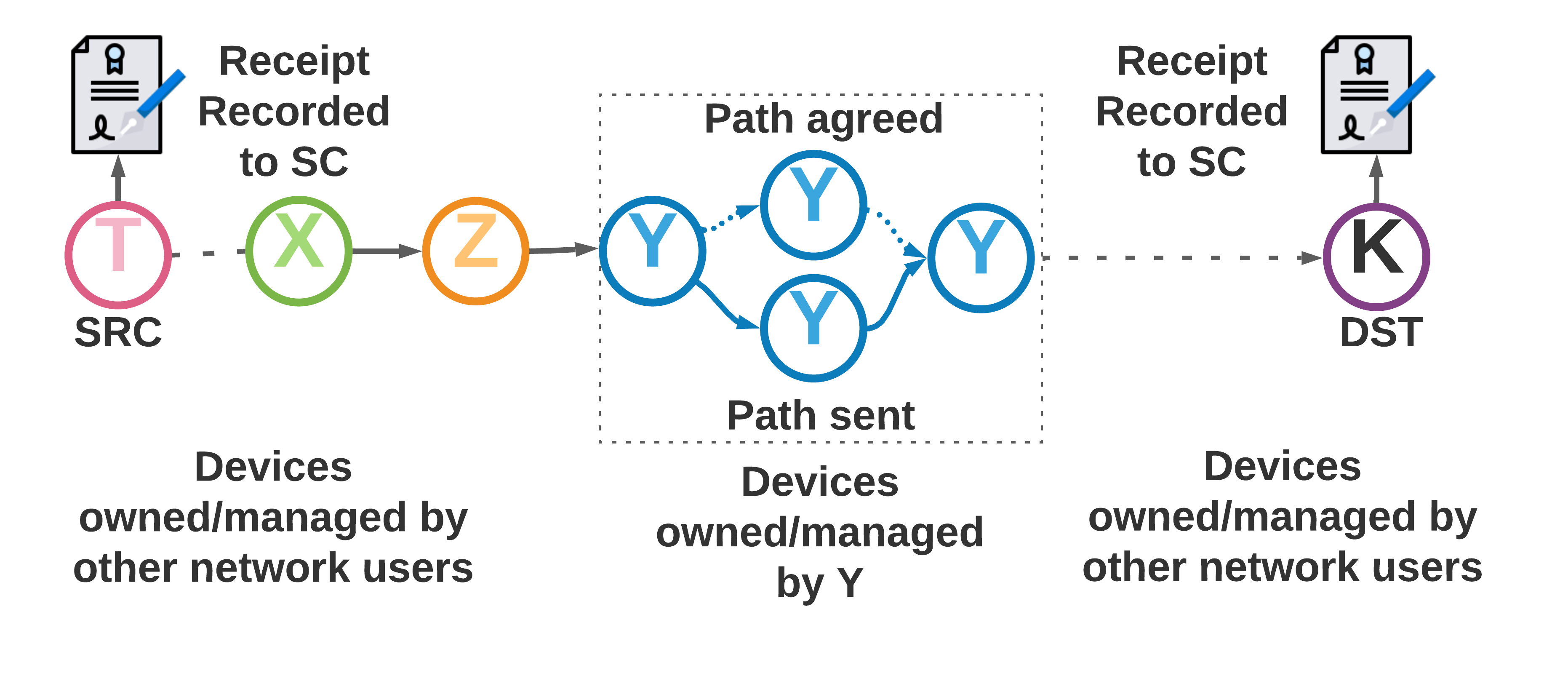}
    \caption{Receipts are recorded at the src (Node T) and dst (Node K) -- the dominant operator Y can allocate devices on a slower and cheaper path rather then the agreed path.}
    \label{fig:threat_model}
\end{figure}

BEAT is an inherently transparent architecture. It is unlikely for network users to misbehave because of two reasons. Firstly, they are allowed in the network with access control mechanisms and are therefore known to the governance and other network users. Secondly, all the receipts are recorded to a PDL, and when data is recorded to the PDL, it can be easily verified later due to its shared replicated record structure. However, there still exists the possibility that any dominant network user, for example, a device owner with a majority of the devices in the network path can cheat -- for example, they can allocate devices to their tenants on a slower and cheaper path instead of the agreed and expensive path without the tenants' knowledge; see Figure~\ref{fig:threat_model}. 


In this section, we propose BEAT's Interrogation protocol which is invoked when any network user~(e.g. tenant) believes that they have not received the agreed service quality. The Interrogation protocol is underpinned by the two key elements of localised record maintenance and  forwarding proof. The combination of these two enables future auditability.
\subsection{Forwarding Proof}

In BEAT, the accountability is managed through the \textit{Receipts}~(Equation~\ref{eq:data}). The receipts are recorded at the source and the destination of the packet. In the event of SLA  violation, device owners/vendors will be required to prove that they have allocated the devices from the source to the destination, as agreed in the SLA~(Figure~\ref{fig:threat_model}). Receipts recording to the PDL through a smart contract at every device is not feasible due to PDL performance considerations. To this end, we propose a \textit{Forwarding Proof} -- a lightweight mechanism for network users to show that they have allocated the SLA-agreed devices throughout the service. The forwarding proofs are stored in local storage and produced \emph{on-demand} to the governance and are not exchanged until a dispute occurs to save the network bandwidth. Because PDLs are inherently transparent and permissioned, we do not need to use any intensive computation techniques to create proofs, and a simple mechanism will suffice. The only objective is to identify the network user (if they have) who has abused the system.



For the $i^{th}$ packet, Node ID is $N$, source ID is $src\_ID_i$, destination ID is $dst\_ID_i$ and timestamp is $tr_i$. Here, source and destination are node's direct neighbours and may not be the final source and destination of the packet. The forwarding proof $P_i$ can be calculated as:

\begin{equation}
    P_i = N + src\_ID_i+dst\_ID_i +tr_i
    \label{eq:receipt}
\end{equation}

Note that receipt timestamp $tr_i$ is the time when the received proof is recorded to the internal storage. This is different from the packet timestamp in Equation~\ref{eq:data} which is the timestamp of the packet received at the device.

\subsection{Localised Record Maintenance}
\label{subsec:local_record_main}
\begin{figure}[!h]
    \centering
    \includegraphics[width=0.40\textwidth]{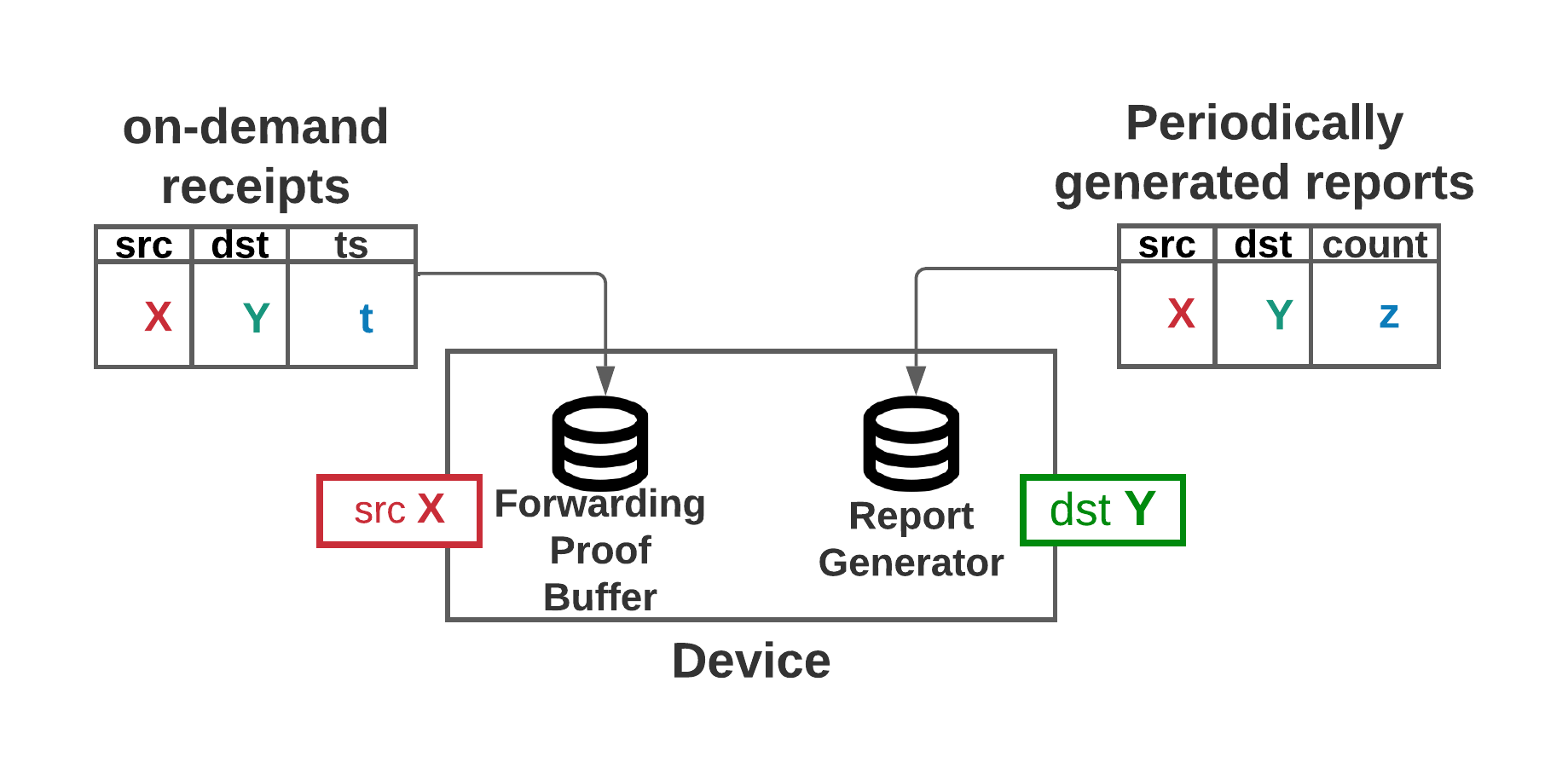}
    \caption{Simplified network device architecture depicting only the two different internal storage BEAT maintains.}
    \label{fig:local_storages}
\end{figure}

BEAT maintains two different types of storage for record keeping; a short-term storage called the \textit{Proof Buffer} and a long-term storage called the \textit{Report Generator} -- Figure~\ref{fig:local_storages}. 

\subsubsection{Proof Buffer}
\label{subsubsec:proof_buffer}
Packet Buffer is a short-term volatile storage that stores the forwarding proof~(Equation~\ref{eq:receipt})~for a governance-defined time called the \textit{Threshold Time}. This threshold time is dependent on the PDL network priorities and available resources. The Proof is kept in a local storage for a threshold time, and after this time elapses it can be overwritten by newer proofs. 

\subsubsection{Report Generator}
For the integrity and compliance monitoring of the PDLs, governance should get periodic reports of the network users. Network devices generate large amounts of data, which makes it very difficult to keep all the data in the device for a long time. The Report Generator is a long-term non-volatile storage that saves the packet count only~(Equation~\ref{eq:report_generator}). For every incoming packet, the software will increment the counter by 1, and at the end of service provisioning, it will send the report to the governance. We are standardising the notion of periodic smart contract reports in ETSI ISG PDL 11~\cite{etsi_pdl}. Given src\_ID and dst\_ID as source and destination identifiers respectively and are node's direct neighbours and may not be the final source and destination. The node $N$ will store record $R$ as.
\begin{equation}
    R =  (src\_ID, dst\_ID,number\_of\_packets)
   \label{eq:report_generator}
   \end{equation}






\subsection{Interrogation Protocol}

If a customer believes that they are not assigned devices as per the agreed SLA and their performance is affected~(e.g. throughput degradation), they can ask the governance to initiate the \emph{ Interrogation Protocol}. 

Recall from Section~\ref{sec:sys_arch2}, that the governance  maintains the record of allocated resources in the SLA for all the resource assignments. Once the dispute is launched, the governance will initiate  the Interrogation Protocol and ask all the devices in the route to send their proof buffers. If the proofs match with the governance records, it means that the devices were assigned honestly, and there was a problem with service quality in the network.

When the governance confirms that all the devices are assigned as per the SLA, further investigations of throughput degradation due to problems such as link failure and packet drop/loss should be carried out. As BEAT is focused fundamentally on resource assignment, service degradation due to other copiously discussed factors such as packet loss and delay~\cite{argyraki2007loss}~\cite{goldberg2008path}, prioritisation attacks~\cite{nikolopoulos2019retroactive} and devices intentionally dropping packet~\cite{zhang2008packet} are out of scope for this work, and we leave these details for  future enhancements.  Nevertheless, the throughput and packet loss per device can still be calculated with the timestamps in the receipts ($ts$) and forwarding proofs ($tr$). 

\begin{figure}[!h]
    \centering
    \includegraphics[width=0.40\textwidth]{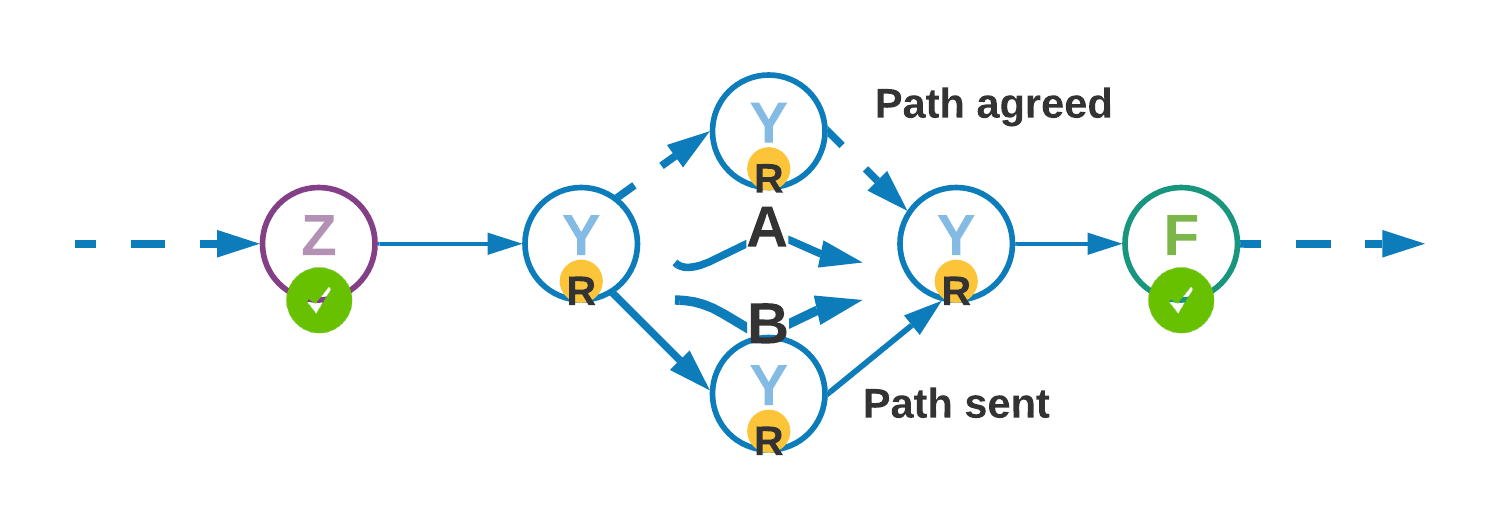}
    \caption{BEAT Threat Model -- the malicious user Y has replicated the proofs to both of their paths to lie to their tenants that they have assigned devices on the promised path. In such a case honest neighbours who also record the source ID can verify the source of the device}
    \label{fig:threat_model_2}
\end{figure}

Network users can still be dishonest, as they can install false replicated forwarding proofs on multiple devices and paths. For instance,  one path A~(the agreed path)  can store the same data as another path B~(path sent); see Figure~\ref{fig:threat_model_2}. Note that from Figure~\ref{fig:threat_model},  forwarding proofs are generated by all devices; therefore, if a network user (e.g., User Y in Figure~\ref{fig:threat_model_2}) is being dishonest, neighbouring devices on the route~(e.g., Devices Z and F) also report their proofs with the corresponding timestamps. By matching their source and destination device identities, it can be verified which device has in fact forwarded the traffic. 

\section{Evaluating BEAT}
\label{sec:evaluation}

The aim of our evaluation is to establish the overheads introduced by BEAT. To this end, we installed PDL nodes on every network device within the GNS3 simulated network infrastructure that formed an end-to-end PDL system. 
\begin{figure}
    \centering
\includegraphics[width=80mm]{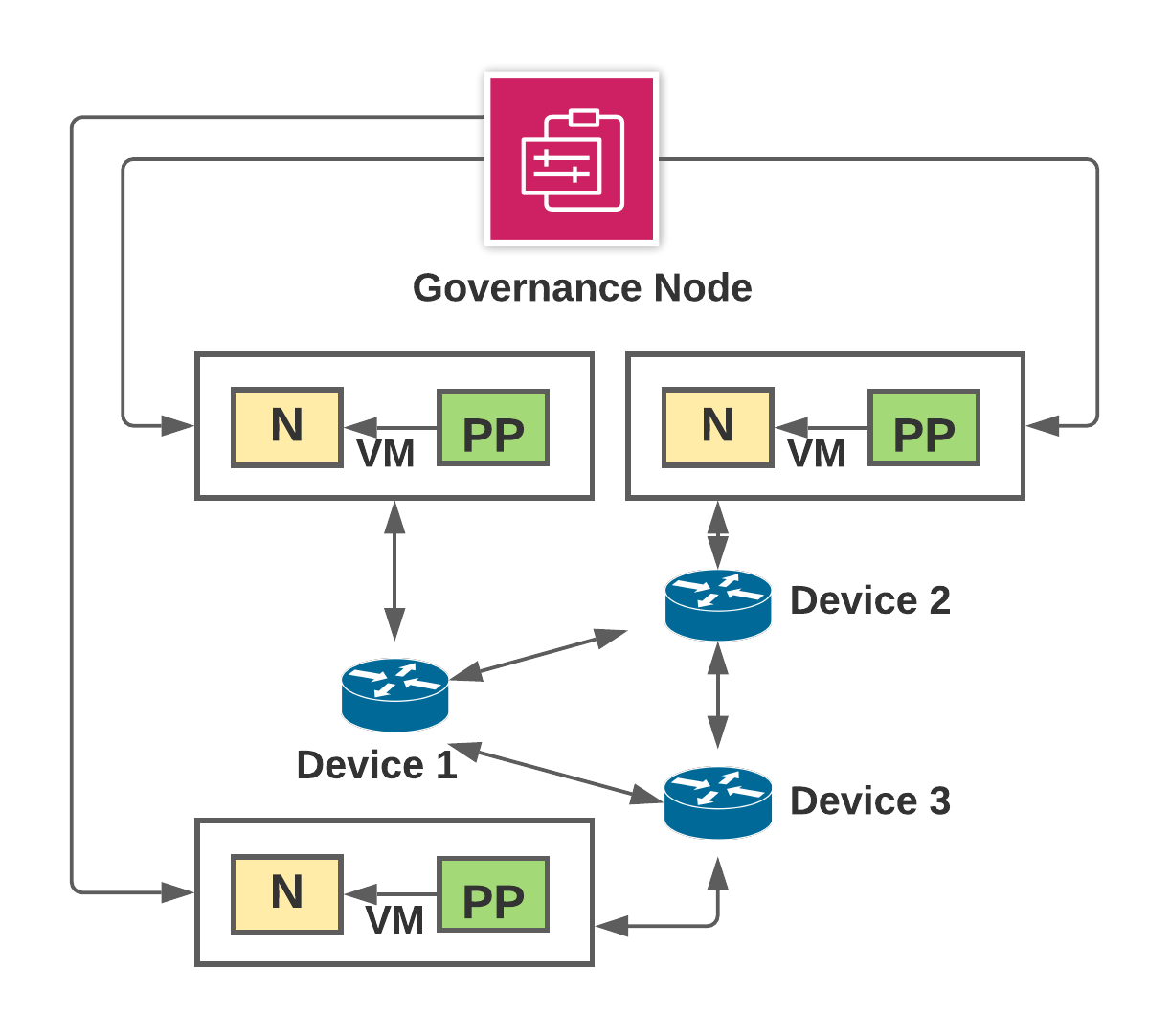}
    \caption{BEAT Simulation Topology -- Each network device has a Packet Processor (PP) and a PDL node (N) installed. Packet Processor sends the captured data to their respective PDL node. Note that the Governance Node does not take part in the consensus and is a logical entity.}
    \label{fig:topo}
\end{figure}
\begin{figure}
    \centering
\includegraphics[width=80mm]{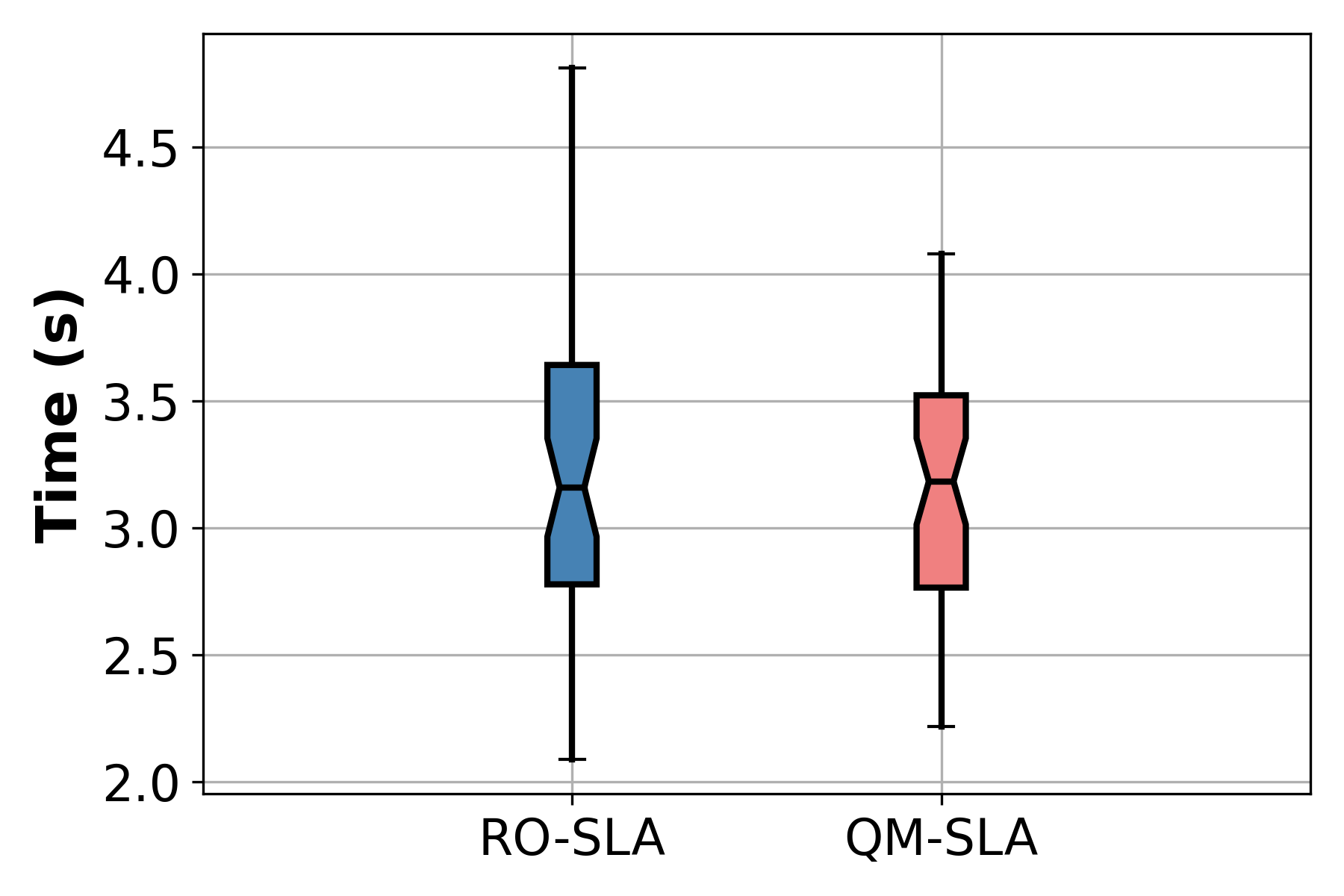}
    \caption{Deployment Latency of a) Resource Orchestration SLA (RO-SLA) and b) QoS Monitoring SLA (QM-SLA) -- Note that SLAs (contracts) are deployed in the PDL in less than 4 seconds.}
    \label{fig:deployment_latency}
\end{figure}


We studied the compatibility between two independent systems, that is, the PDLs and the network infrastructure. We evaluated the viability of our proposal with a simple but similar real-world network topology~(Figure~\ref{fig:topo})  with three network devices. Each device runs a PDL node and is connected to the governance node.

The idea of this study was to enable accountability with PDLs at the network layer - therefore, we installed one Ethereum node on each of the three edge routers. We advocated the use of permissioned distributed ledger -- hence we adopted Ethereum's permissioned version with Proof-of-Authority~(PoA) consensus protocol; more specifically, the ``Clique'' implementation of PoA which has higher throughput and lower latency than traditional Proof-of-Work protocol~\cite{singh2019managing} and blocks can be generated at a user-defined time interval. Clique also outperforms its other PoA counterparts and requires fewer message exchanges and hence delivers high throughput as compared to other permissioned flavours of Ethereum such as Aura~\cite{de2018pbft}. 

We set up our simulations on an Intel Core (dual core with two logical processors) i7 CPU at 2.70 GHz with 16 GB RAM; with GNS3 network simulator.  The blocks for the PDL are generated at 15 seconds intervals. The smart contract to record the usage is coded in Solidity and installed on the ledger.~Two different SQLite databases are maintained to evaluate the Orchestration Process;~see Section~\ref{fig:orch_process}.

\subsection{SLA Deployment Latency}
The focus of this study is on accountable and transparent SLAs and we achieve this through smart contracts. In BEAT, the governance of the PDL is responsible for the deployment of SLA. There are two different types of SLAs are involved in BEAT, 1) Resource Orchestration SLA (RO-SLA), that is, SLAs that allocate the resources, and 2) QoS monitoring SLA (QM-SLA).

\paragraph{Resource Orchestration SLA (RO-SLA)}
The RO-SLA as smart contracts are deployed by the Orchestration Manager. In our simulation the resource orchestration SLA took a minimal time of mean  3.98 seconds;~see Figure~\ref{fig:deployment_latency}.

\paragraph{QoS Monitoring SLA (QM-SLA)}
In our simulation, the installation took mean 3.11 seconds; see Figure~\ref{fig:deployment_latency}.  Typically smart contracts are installed once and can be executed several times as required. Yet, this is dependent on the number of simultaneous transactions allowed by the PDL. 

\subsection{Orchestration Process}

Three key elements for the Orchestration Process~(Figure~\ref{fig:orch_process}) are 1) Access control verification, 2) Resource availability verification, and 3) Execution of the RO-SLA.

\begin{figure}[!h]
    \centering
    \includegraphics[width=80mm]{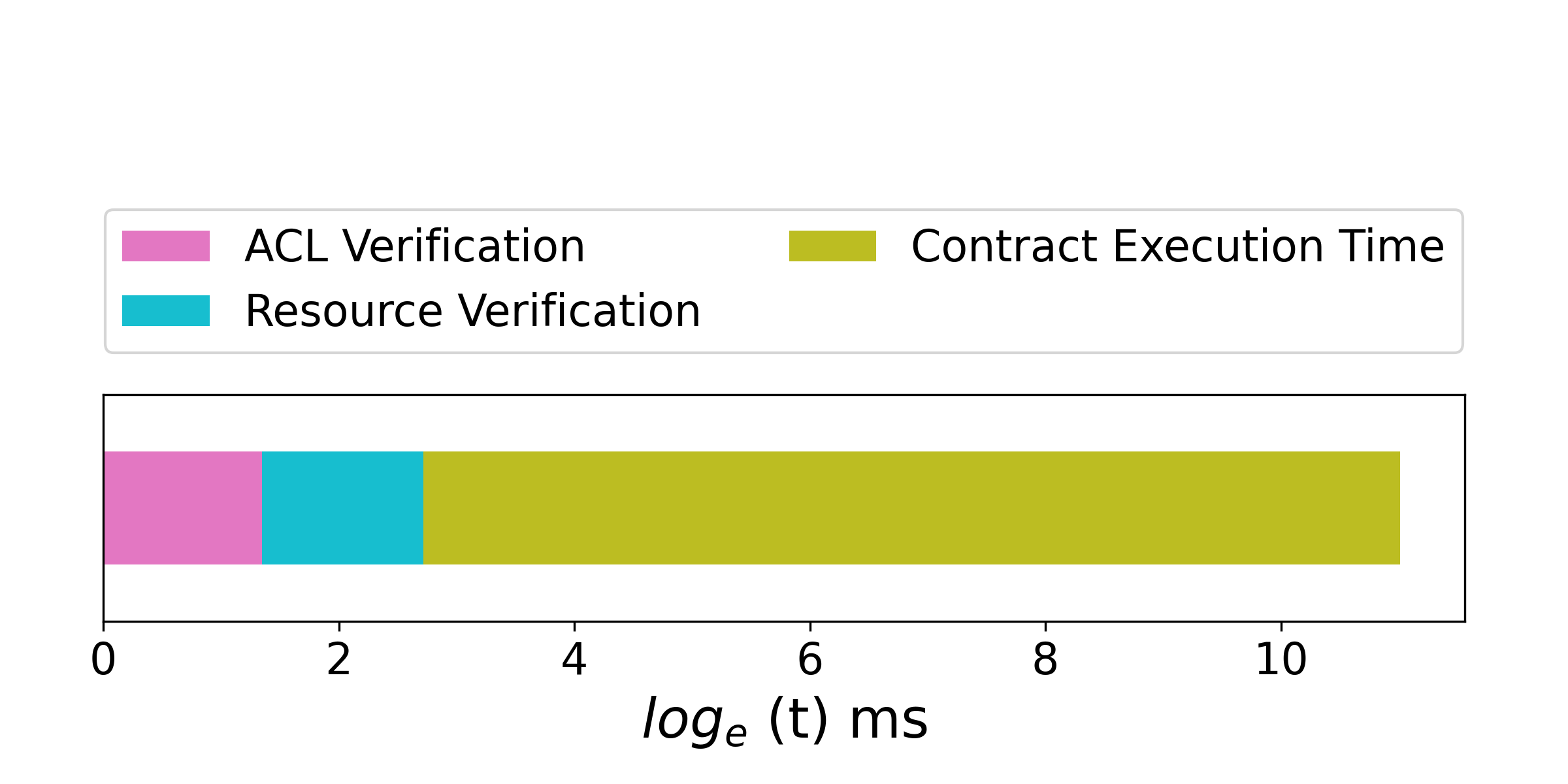}
    \caption{Segregated Orchestration Process overheads -- Note that the overall orchestration process took mean 4 seconds}
    \label{fig:orch_time}
\end{figure}

The orchestration process starts with a resource request to the Orchestration Manger. The Orchestration Manager forwards this request to the access control entity which manages the access control list in a database. As shown in Figure~\ref{fig:orch_time}, the access verified in  3.85 ms. The Orchestration Manager then checks for resource availability from an internal database which took an average of 3.93 ms. If the resources are available the RO-SLA (i.e., smart contract) executes, which took an average of 3.73 seconds; see Figure~\ref{fig:orch_time}.

\begin{figure}[!h]
    \centering
\includegraphics[width=80mm]{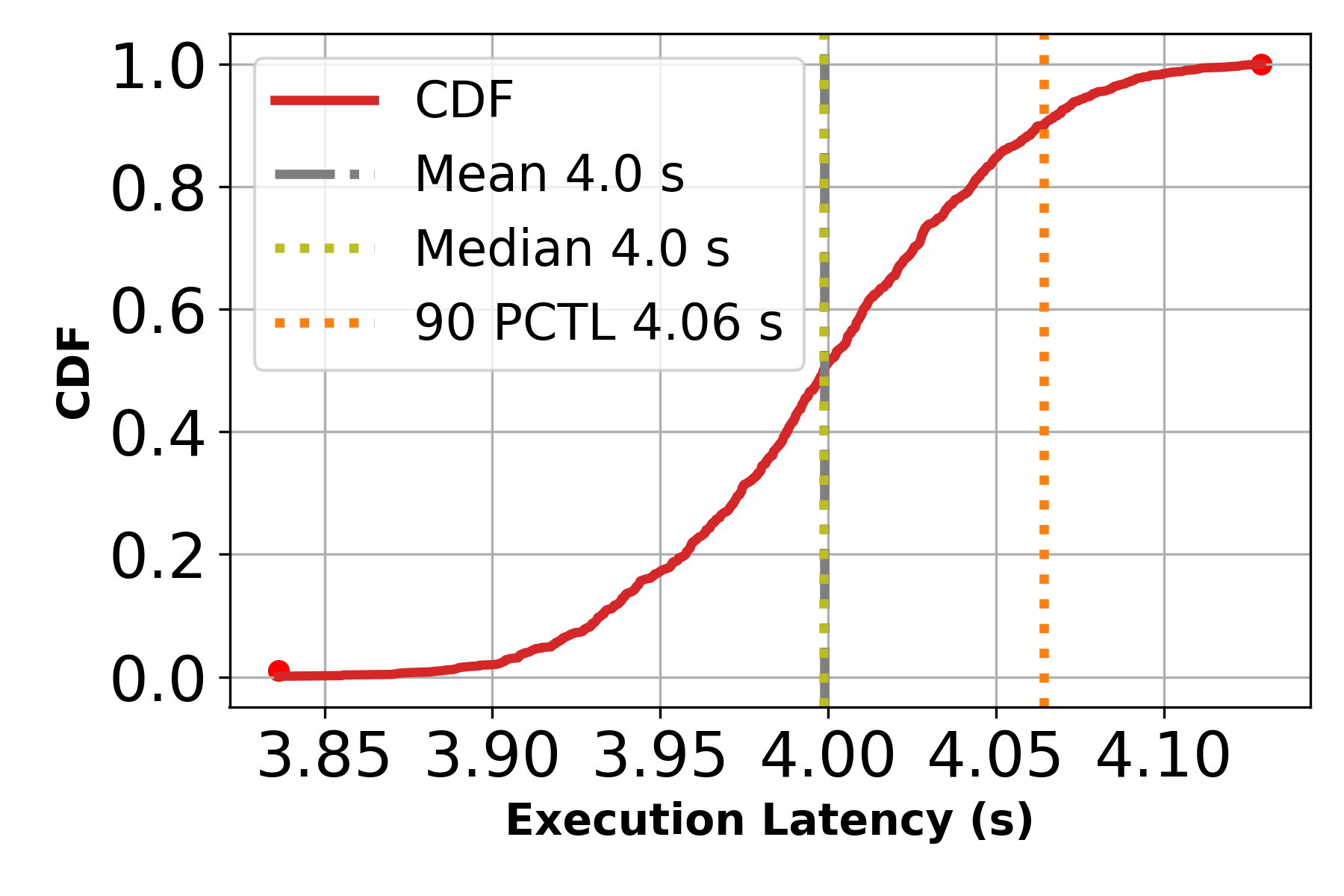}
    \caption{Orchestration Process -- Note that the total orchestration process took an order of just seconds. }
    \label{fig:orch_overall}
\end{figure}

\emph{In summary, the overall orchestration process took around  4 seconds on average; Figure~\ref{fig:orch_overall}}.

\begin{figure*}[!h]
\centering
\begin{subfigure}{0.3\textwidth}
  \includegraphics[width = \textwidth]{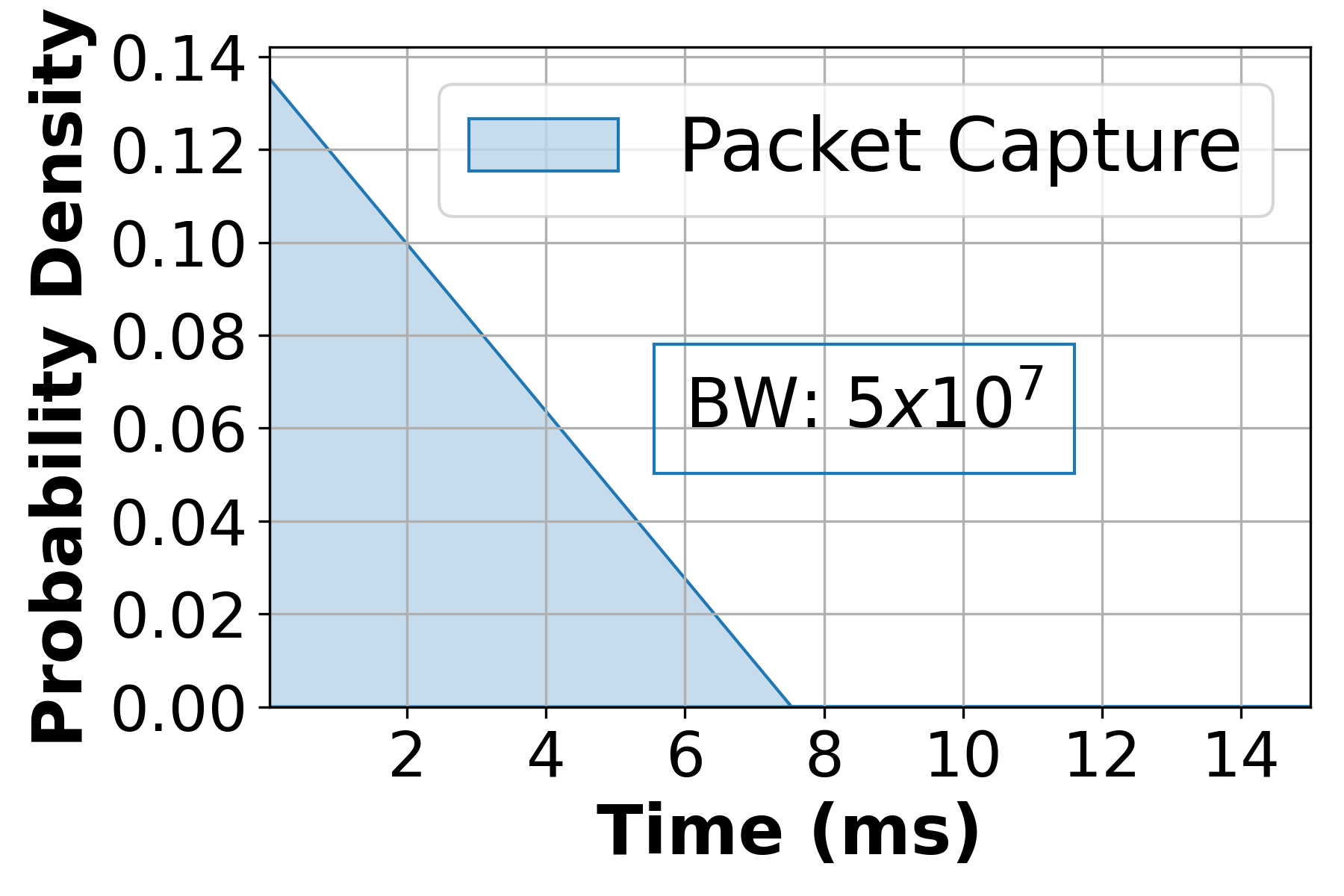}
  \caption{KDE plot for packet capture}
  \label{fig:(a)pc}
\end{subfigure}\hfill
\begin{subfigure}{0.3\textwidth}
  \includegraphics[width = \textwidth]{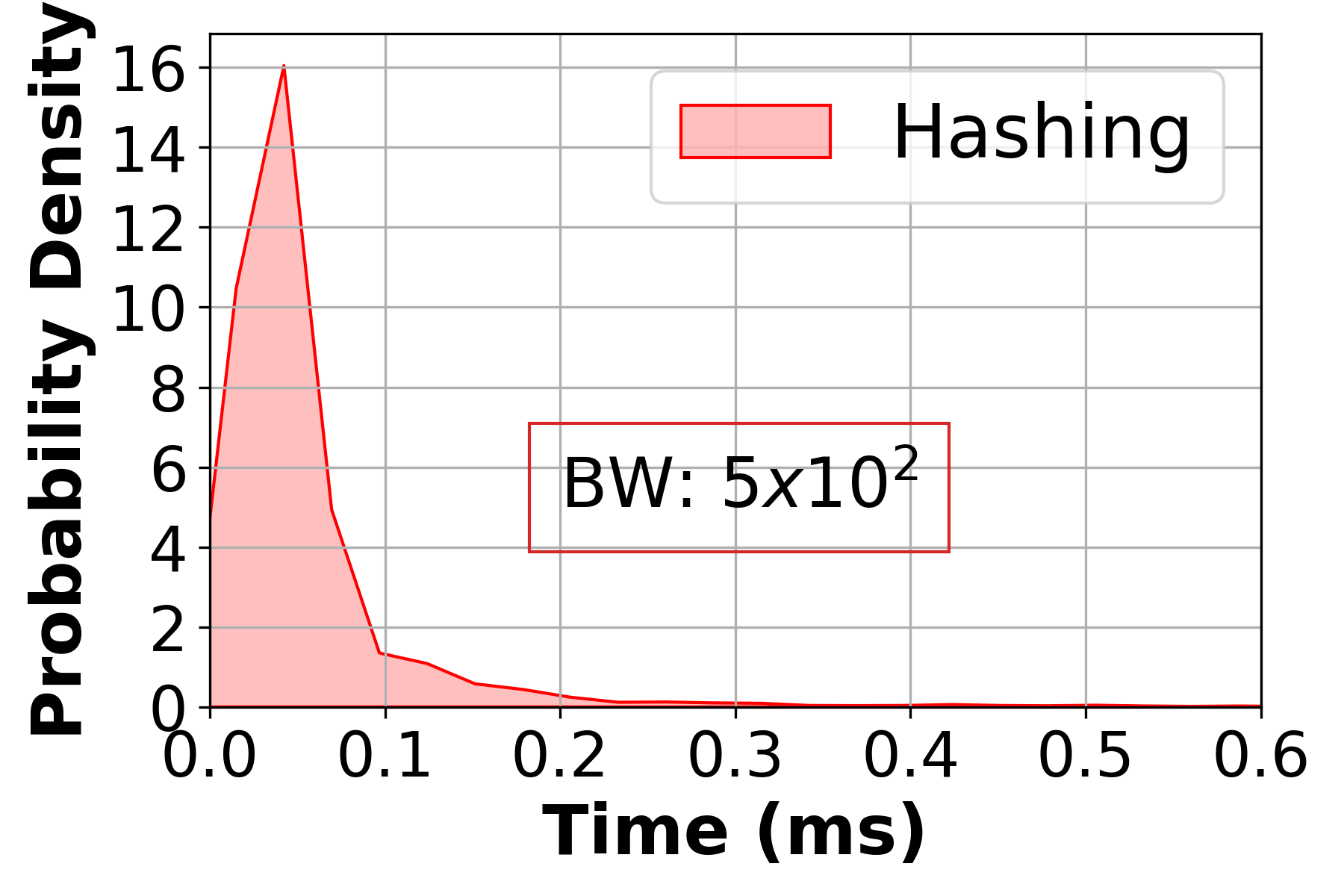}
  \caption{KDE plot for hashing}
  \label{fig:(b)hashing}
\end{subfigure}\hfill
\begin{subfigure}{0.3\textwidth}
  \includegraphics[width = \textwidth]{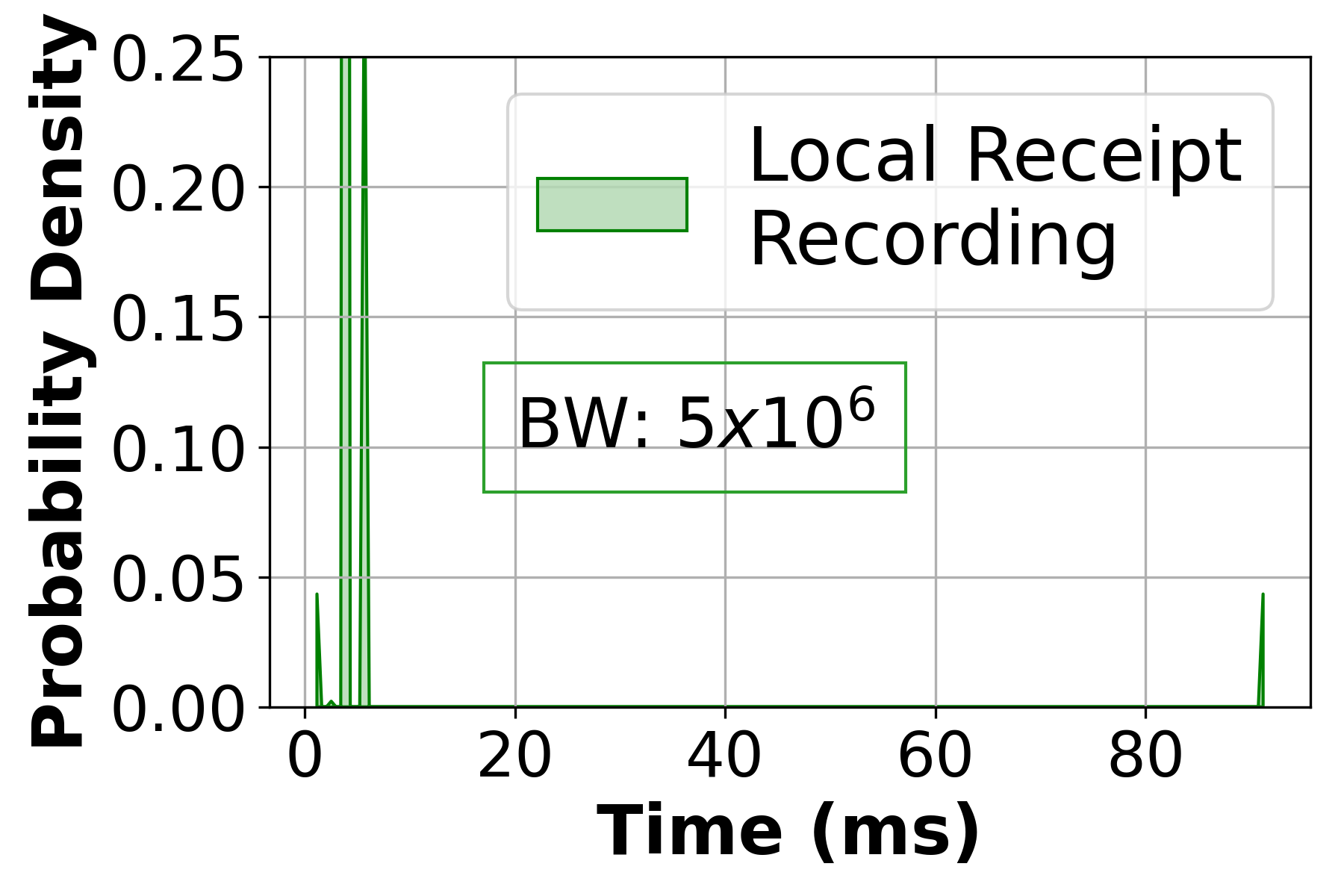}
  \caption{KDE plot for receipt recording}
  \label{fig:(c)db}
\end{subfigure}
\bigskip
\begin{subfigure}{0.3\textwidth}
  \includegraphics[width = \textwidth]{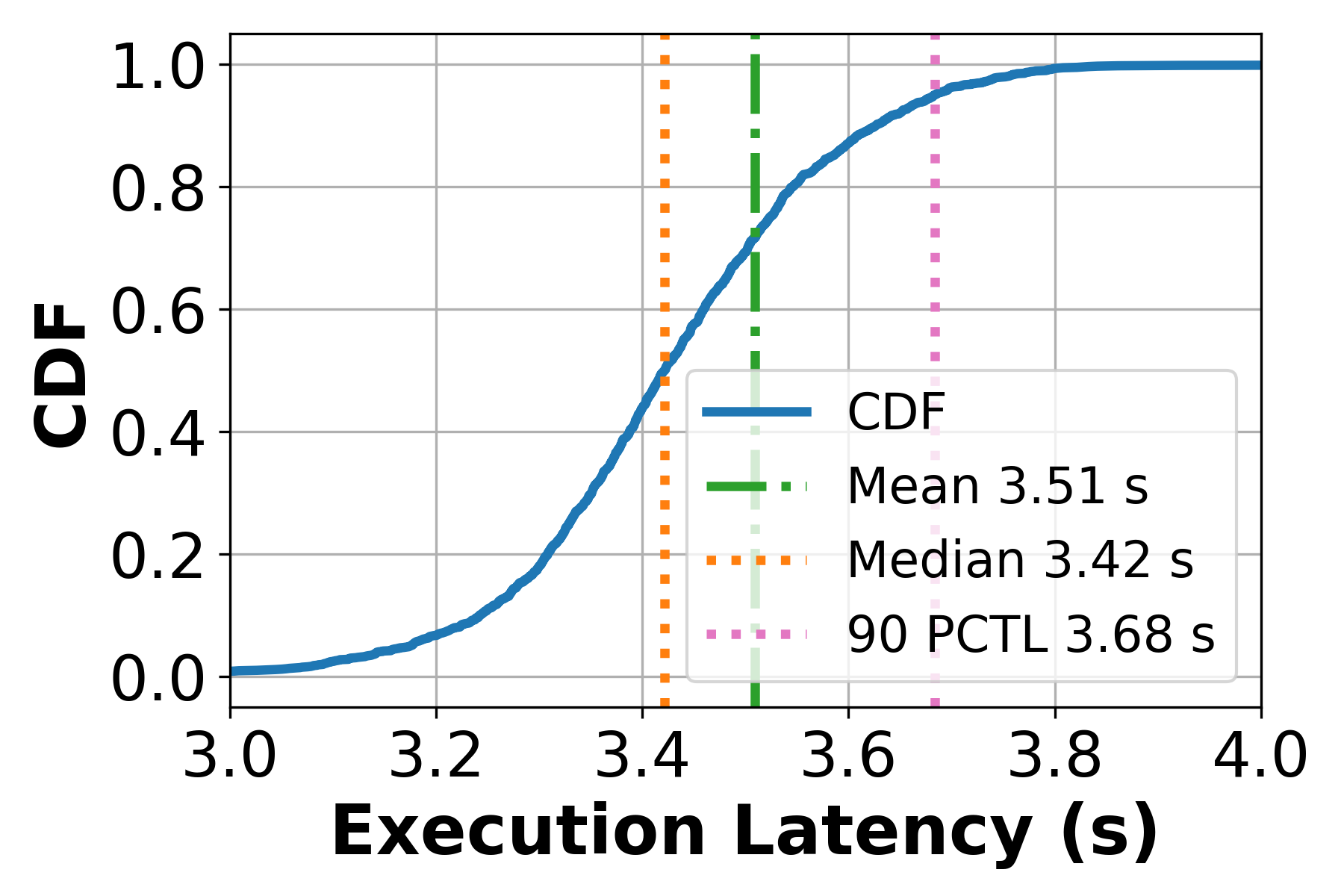}
  \caption{CDF of SLA execution}
  \label{fig:(d)contract_execution}
\end{subfigure}\hfill
\begin{subfigure}{0.3\textwidth}
  \includegraphics[width = \textwidth]{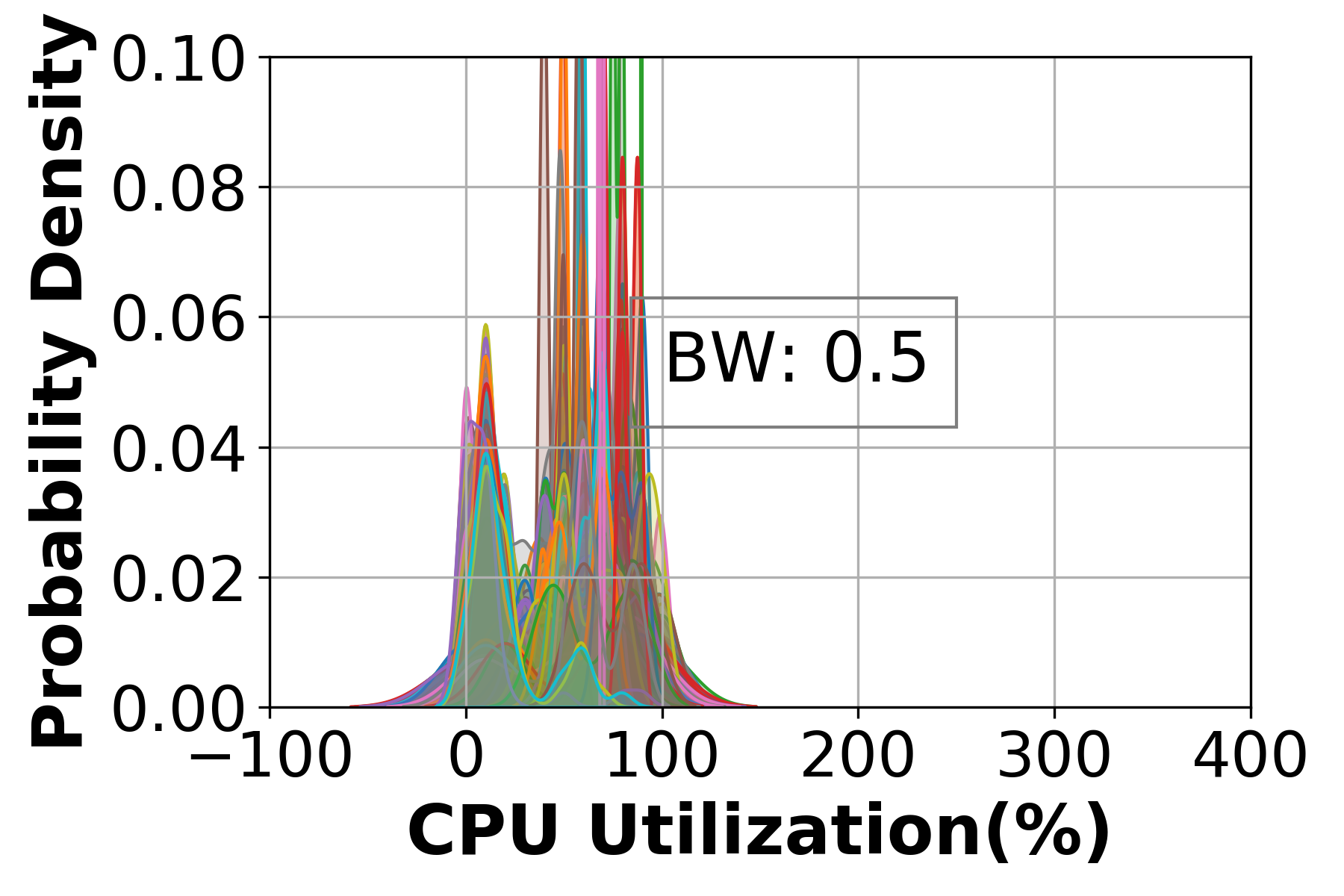}
  \caption{KDE plot for CPU Utilization}
  \label{fig:(e)cpu_utilization}
\end{subfigure}\hfill
\begin{subfigure}{0.3\textwidth}
  \includegraphics[width = \textwidth]{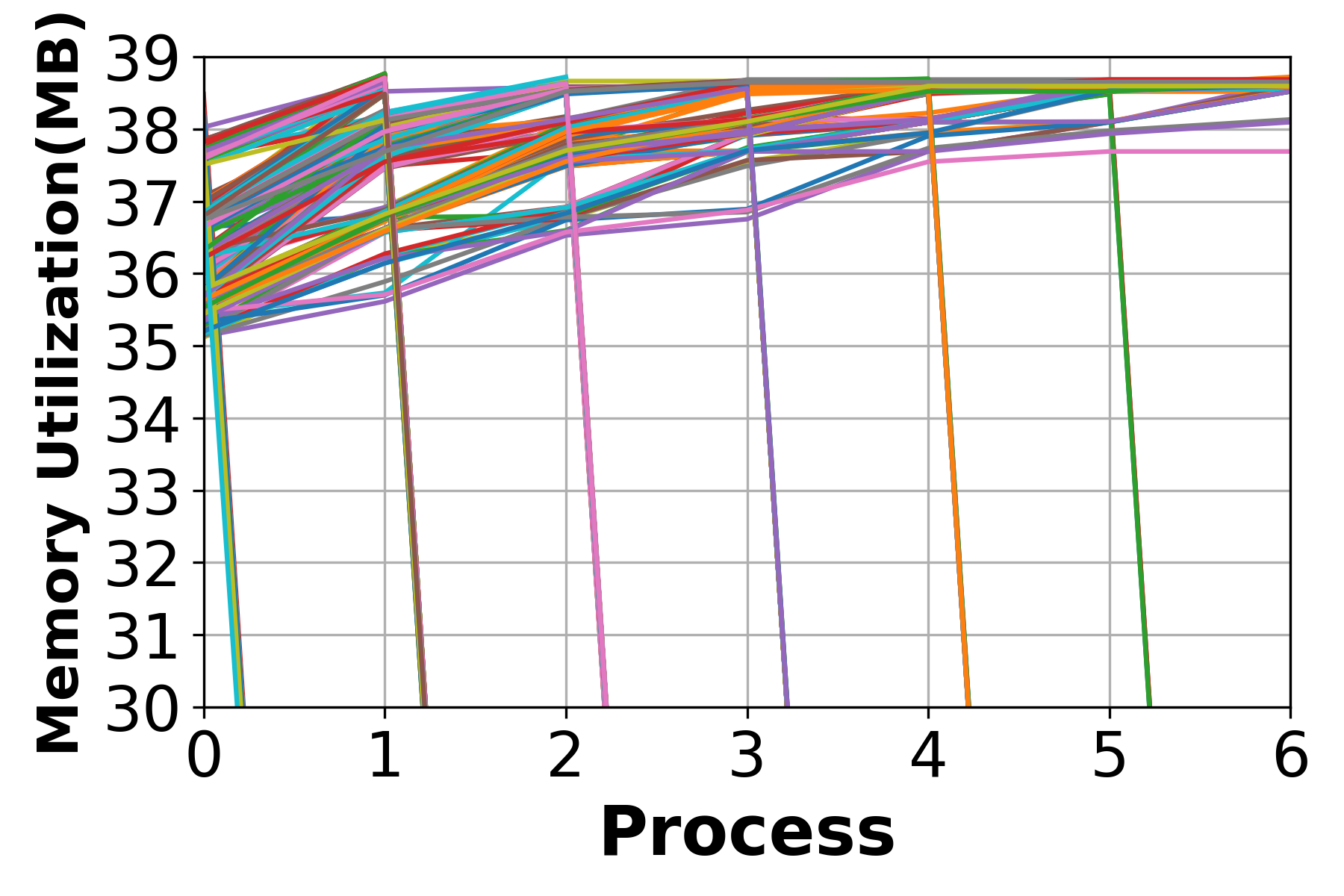}
  \caption{Memory Utilization}
  \label{fig:(f)memory_utilization}
\end{subfigure}

\caption{Packet Processor Overheads -- a) Packet capture, b) Hashing, c) Receipt recording time, d) Contract (smart contract) execution time, e) CPU Utilisation, and f) Memory utilisation are shown. Our results show that, altogether, packet processing and SLA recording adds less than 5 seconds of overhead. Bandwidth (BW) is adjusted in Kernel Density Plots (KDE) for packet capture(a), hashing(b), receipt recording(c) and CPU utilisation (e) to improve the readability and data visualisation. }
\label{fig:homodynReconstrComp}

\end{figure*}
\subsection{Packet Processor}
The next step is to start the packet forwarding process. The \textit{packet processor} shown in Figure~\ref{fig:packet_processor} is coded with Python 3  and installed on a virtual machine adjacent to PDL node.~We sent almost 4000 flows in our topology. The Packet Processor captures and processes the flow data and records to the PDL. In our simulations, it took an average of 1.02 ms~(Figure~\ref{fig:(a)pc}) to capture the packet and extract the relevant data (i.e., \emph{node ID}, \emph{source IP address}, \emph{destination IP address}, and \emph{timestamp}). The Packet Processor then, hashes this data (Equation~\ref{eq:data}) and records it to the PDL through smart contract execution. We advocate the use of a pluggable hashing algorithm -- that is, the governance of the PDL has the liberty to choose their preferred algorithm based on their priorities, such as the availability of computational power.~In this study, however, we chose the 32-byte version of SHA3, that is, SHA3-256. In our experiments, it took an average of 0.06 ms to hash the 48  to 56 bytes processed data; see~Figure~\ref{fig:(b)hashing}.

Recall from Section~\ref{subsec:local_record_main}, that BEAT also maintains local storage on every node to maintain off-chain records for future auditability. In our simulations it took just an order of milliseconds~(specifically, 6.15 ms mean) to record the  forwarding proofs~(Equation~\ref{eq:receipt}) to an internal storage (Figure~\ref{fig:(c)db}).

The receipts are recorded to the PDL at the first and last devices only. They are recorded through the execution of a pre-installed smart contract and took an average of 3.51 seconds in our simulations (Figure~\ref{fig:(d)contract_execution}).  

Indeed, like any other software running PDL requires additional computational resources. Note that, our setup is in a simulation environment where network switches are, infact, containers. To measure the memory and CPU utilisation of BEAT, we randomly chose 500 records. CPU utilisation is measured every 0.1 seconds during the packet capture process, and the results are shown in Figure~\ref{fig:(e)cpu_utilization}. The CPU  utilization has a maximum median 89.45\% and the maximum value is 109.20\%\footnote{the maximum CPU utilization on a dual-core CPU can go upto 200\%} across all 4000 flows.

The memory utilisation is shown in Figure~\ref{fig:(f)memory_utilization} -- memory is measured during the packet processing time and we note that in some cases the memory utilisation ends before the rest of them. This implies that the contract is executed faster in some cases. Note that, most of the data is in between 35 - 39 MB and the maximum memory utilization is 39 MB.  In other words, \emph{BEAT adds a negligible overhead to the system. Note that our simulations were performed on a local laptop. All the devices were infact virtualised containers and shared the same hardware. In a production environment with specialised network devices, we anticipate far better performance.}

In a private conversation with a major vendor it is confirmed to us that, protocols proposed by BEAT are realistic. It was also discussed that it is realistic to expect a better performance in the production environment due to state-of-the-art equipment with higher computational power and sufficient memory storage. 





\section{Considerations}
\label{sec:considerations}

Designing a system for 6G and the PDLs is not trivial. We advocate PDLs for BEAT, in permissioned distributed ledgers, the governance oversees the network operations such as the allowed number of users and participants' misbehaviour. Most of the challenges can be resolved with the compliance strategies implemented by the governance. However, some considerations, such as transaction throughput and scalability, need attention.~In this section, we discuss the considerations related to the BEAT.
\subsection{Inherent Networks Attacks}
\subsubsection{Intra-PDL Denial-Of-Service~(DoS)}
Every PDL allows a particular number of Transactions Per 
Second~(TPS), which are generally higher in PDLs (e.g., 20,000 TPS~\cite{gorenflo2020fastfabric}) than in permissionless ledgers~(e.g., Ethereum mainnet which has a TPS of $\approx$ 15). This means that if many devices send data simultaneously, it can cause congestion at the ledger or DoS for incoming transactions. Therefore, whilst designing a PDL network, it is important to adopt a PDL-type (e.g., Hyperledger Fabric), which can cope with the network's requirements. 
In this work, we advocated the use of PDLs, due to their high transaction throughput and stringent access control mechanisms only the number of users that the PDL system can cope with, will be allowed by the governance. Also note from Section~\ref{subsec:orch_layer}, the Orchestration Layer maintains the record of available resources and active users. The access can be denied when the network reaches a certain threshold.
\subsubsection{Denial of Capability (DoC) Attack}

BEAT's permissioned nature makes inter-PDL DoS attacks unviable, yet, general DoS attacks (e.g., DoS due to outside network users) are still possible. That is, malicious users can send an inflated number of requests to the Orchestration Manager~(Section~\ref{sec:sys_arch2}), causing congestion. Several solutions to combat DoS exist and can be used to mitigate the problem. However, in this work, we support capability-based solutions~\cite{anderson2004preventing}. The reason to choose capability-based solutions is that, in such systems, the requesting traffic is limited to a small channel, and the rest of the bandwidth is dedicated to authorised traffic for general usage. During the requesting period, users will be assigned a token that allows them to access the network on the main channel. Based on criteria and user behaviour, a user can be assigned a token for a longer time duration. This is a viable approach in BEAT because BEAT will need to communicate with its client only at the setup time. This will involve a small number of communications between a client and the network; also, the clients are likely to be legitimate and safe to be assigned tokens in longer terms. Limiting the connection requests to a dedicated channel will allow only the authorised traffic (from clients such as OTT) to further communicate with the orchestration manager. 

However, the requesting channel is still open to unauthorised users to send requests. A large number of connection requests flooding the request channel can cause congestion at the requesting channel. Consequently, legitimate users are unable to send the connection request to the network. This is identified as ``Denial-of-Capability Attack'' by Argyraki et al.~\cite{argyraki2005network} and they argued against their viability in combating the DoS attack in~\cite{argyraki2005network}. However, other research works argue Argraki et al.'s argument and propose enhancements to solve the problem at the requesting channel, such as stateless-filtering \cite{yaar2004siff} and puzzles~\cite{parno2007portcullis}. BEAT can implement these solutions to solve the problem of DoC attacks. The details on DoS and DoC is beyond the scope of this work.

Note that, in this work, we argued on and supported capability-based solutions to combat DoS attacks for BEAT. However, a number of other solutions are proposed to mitigate DoS attacks, such as \cite{argyraki2005active} and \cite{yang2005limiting} and can be adopted and used to mitigate DoS attacks in BEAT. 
\subsubsection{Malicious Devices}
All the devices record the data to the PDL; the data is replicated across the network nodes~(i.e., network devices). If a malicious device starts sending false and irrelevant data to the PDL, the system can get overwhelmed by the number of requests and increase the data sizes inside the routers exponentially. 

To combat this problem, BEAT maintains a node ID for all the nodes in the network (Section~\ref{sec:sys_arch2}). In case of misbehaviour, with nodes IDs the governance can take appropriate actions and block future access of such nodes. Indeed, such an enforcement requires standardisation, and the authors of this paper have proposed standardised compliance strategies to manage node misbehaviour in ETSI Group Specification PDL 11~\cite{etsi_pdl}. 

\subsection{Other Threats}
\paragraph{Integrity of Data}
Smart contracts do not have any built-in means to verify the integrity of the data. Hence, it is vital to ensure that the data recorded by network devices is valid. 
In BEAT, TEE ensures that the correct data is recorded to the PDL. However, the network device feeds the data to the packet processor and if this device is malicious, it will provide false information to the PDL.

This is one of the reasons we advocated the use of governance-controlled \emph{Permissioned} Distributed Ledgers. In a PDL, the members are allowed with access control mechanisms and affected parties can report such misbehaviour to the governance, which can take subsequently disciplinary actions and \emph{blacklist} the node and may impose penalties~\cite{etsi_pdl}.

\paragraph{Colluding MSPs}

In a Blockchain-enabled architecture like BEAT, the network users can collude with each other. In such a case, dominant network users can behave maliciously towards other tenants such as, rejecting their transactions. To solve such problems, a regulatory authority~(e.g. Ofcom in the UK and Federal Communications Commission (FCC) in the US) can also be a part of the PDL network governance. Note that, the role of the regulatory authority is as an observation entity only and it should be contacted only in the event of disputes. The regulatory authority neither takes part in consensus nor controls any device. 


To add additional layer of security it is also possible to record the complete path of the packet with BEAT and record the data on every device if scalability and congestion of the PDL are not a concern.
\paragraph{Waiting Times}
Recall that the Orchestration Manager allocates the resources considering the available capacity of the network. Therefore, it is likely that some users will have to wait to get hold of the resources. BEAT is an accountability focused architecture, that is, adherent to the SLA.~Controlled resource allocation ensures that the network is not overwhelmed by the service requests and that users receive SLA-promised services.

\section{Future Work}
\label{sec:future_work}
\subsection{Artificial Intelligence-Enabled BEAT}

Blockchain and AI are two key enablers of 6G. BEAT is inherently an AI-enabled architecture. All of the functions which are manual at this stage, can be in principle automated through AI.

\subsubsection{Automated SLA Violation Reporting}
At this stage, BEAT relies on the PDL members to report SLA violations. Using AI algorithms such as logic regression and Hidden Markov Model, the anomalies in the network can be detected~\cite{li2017intelligent} and reported automatically. In AI-enabled BEAT, SLA violations will be reported without any human intervention.

\subsubsection{Automated Governance}
In this work, we propose that governance to be formed through representatives from all the members of the PDL. However, governance can also be an automated functionality. i.e.,  software code, programmed and approved by all PDL members,  which takes decisions instead of human. 

Artificially intelligent governance, indeed, can play an important role in future corporate governance~\cite{hilb2020toward}. Through AI, the governance model can be trained to make operational decisions such as disputes.

\subsubsection{Intelligent Smart Contracts}
At this stage, smart contracts in BEAT are pre-programmed and installed on the PDL by the governance;~Section~\ref{sec:sys_arch2}. In the future, we aim to install AI-based Intelligent Smart contracts. These smart contracts can implement smart metering and automated SLA negotiation.

\subsection{Telco-Distributed Ledger}
Most of today's available PDLs~(e.g., Hyperledger Fabric and Corda) can be used atop BEAT. However, we aim to design a telco-focused permissioned ledger:\emph{``T-Chain''}. In infrastructure sharing architectures like BEAT, the network users are typically known to each other. The primary requirement is an immutable, transparent and automated  contract management system. In T-Chain, the consensus algorithm is lightweight, and all the transactions do not need to be approved by all the members, but approval from the governance node is sufficient. 


\section{Conclusion}
\label{sec:conclusion}

In this work, we have presented ``BEAT'', a PDL focused automated, transparent,  accountable network sharing architecture operating at the network layer. In future generation networks, operators need to work collaboratively to broaden their coverage area and cope with the ever-increasing demand for network services. A key enabler for viable network sharing is accountability and transparency at every layer of the network sharing architecture. In this work, we focused on the network layer. 

In BEAT, the SLAs between the network users are recorded as smart contracts in a PDL. We introduced a layered architecture in which governance of the PDL manages and maintains the network resources with stringent access control and network management strategies.~Yet, some of the network users can still misbehave, for example, by allocating resources on a cheaper and slower path instead of the agreed path. To this end, the BEAT maintains record in all of the network devices. This record/receipt is presented by the devices only in the situations of dispute, enabling a lightweight audit mechanism.

BEAT adds a negligible overhead  to the system, when considering the time and cost required for the negotiation of SLAs for infrastructure sharing. We believe that our system provides a faster and more seamlessly automated solution and is the future of infrastructure sharing.


In conclusion,  it is our hope that this work  marks  the  inception of a new era of network sharing in which competing stakeholders can work efficiently and transparently to achieve a scalable and open network.

\bibliographystyle{IEEEtran}
\bibliography{bib/acm}

\vspace{0.1pt}
\begin{IEEEbiography}[{\includegraphics[width=1.5in,height=1.25in,clip,keepaspectratio]{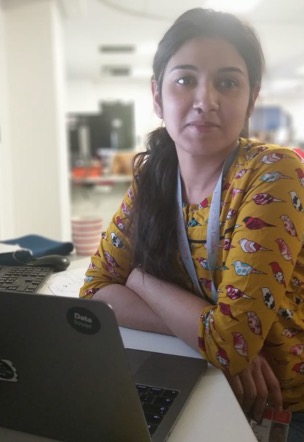}}]{Tooba Faisal} is a PhD student at King's College London (KCL) working on Telco-blockchain. Her current research interests are designing secure smart contracts, Distributed Ledger Technology and SLAs. She also has a Master of Research (MRes) in Security Science from University College London, a Master of Science (MS) in Telecommunication and Networks and a Bachelor of Science degree in Computer Engineering from Bahria University, Karachi, Pakistan. She is a KCL's delegate in the ETSI Industries Specifications Group on Permissioned Distributed Ledgers and rapporteur of smart contract Group Report and Specifications. 
\end{IEEEbiography}
\vspace{-1cm}
\begin{IEEEbiography}[{\includegraphics[width=1.5in,height=1.25in,clip,keepaspectratio]{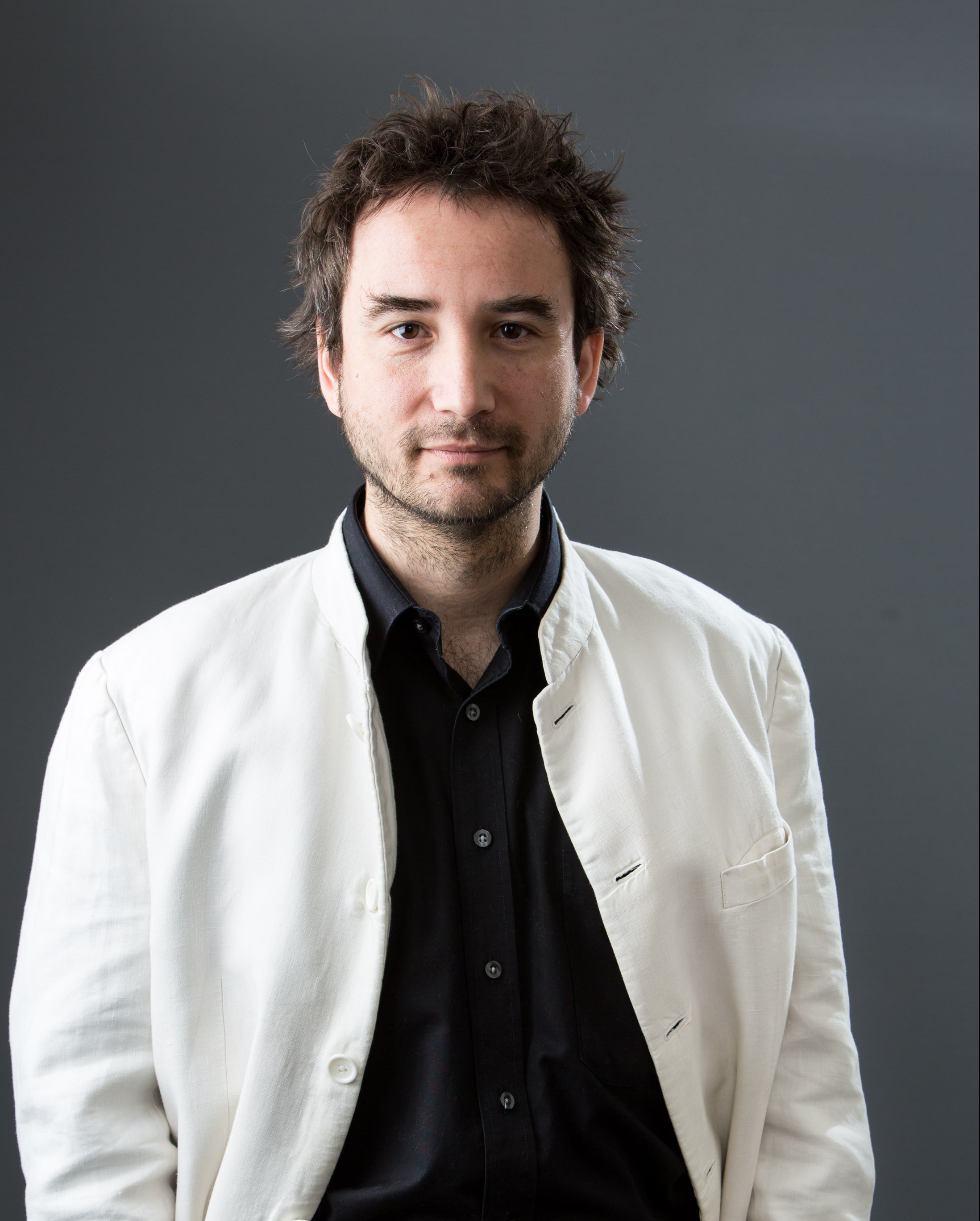}}]{Mischa Dohler} now Chief Architect at Ericsson Inc. He was Professor at King’s from 2013-2021, and the Director of the Centre for Telecommunications Research from 2014-2018. He also worked as a Senior Researcher at Orange/France Telecom from 2005-2008. He is a Fellow of the IEEE, the Royal Academy of Engineering, the Royal Society of Arts (RSA), the Institution of Engineering and Technology (IET); and a Distinguished Member of Harvard Square Leaders Excellence. He sits on the Spectrum Advisory Board of Ofcom, and acts as policy advisor on issues related to digital, skills and education. 
\end{IEEEbiography}
\begin{IEEEbiography}[{\includegraphics[width=1in,height=1.25in,clip,keepaspectratio]{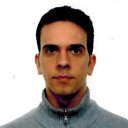}}]{Simone Mangiante} received his Ph.D. in computer networks in 2013 from the University of Genoa, Italy, working on carrier Ethernet management using the SDN paradigm. He then spent three years with Dell EMC in Ireland as a senior research scientist, where he managed European projects focusing on SDN and network transport. He led the design and deployment of a virtualized industrial IoT testbed and contributed to several H2020 EU proposals. He is currently a research and standards specialist in Vodafone Group, United Kingdom. His main research interests are edge computing, distributed cloud architecture, and NFV. 
\end{IEEEbiography}
\begin{IEEEbiography}[{\includegraphics[width=1in,height=1.25in,clip,keepaspectratio]{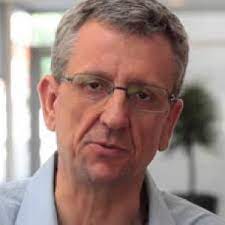}}]{Deigo R. Lopez}
 joined Telefonica I+D in 2011 as a Senior Technology Expert on network middleware and services. He is currently in charge of the Technology Exploration activities within the GCTO Unit of Telefónica I+D. Before joining Telefónica he spent some years in the academic sector, dedicated to research on network service abstractions and thedevelopment of APIs based on them. During this period he was appointed as member of the High Level Expert Group on Scientific Data Infrastructures by the European Commission.~Diego is currently focused on identifying and evaluating new opportunities in technologies applicable to network infrastructures, and the coordination of national and international collaboration activities. 
\end{IEEEbiography}
\end{document}